\documentclass[10pt,journal,compsoc]{IEEEtran}
\usepackage{cite}
\usepackage{amsmath,amssymb,amsfonts}

\usepackage[usenames,dvipsnames]{xcolor}
\usepackage{subcaption}
\usepackage{booktabs} 
\usepackage{array}
\newcolumntype{x}[1]{>{\centering\arraybackslash\hspace{0pt}}p{#1}}

\usepackage{extarrows}

\usepackage{algorithm}
\usepackage[noend]{algpseudocode}
\usepackage{amsmath}
\usepackage{soul}
\usepackage{graphicx}
\usepackage{textcomp}
\usepackage{pgfplots}
\usepackage{soul}

\pgfplotsset{width=9cm,compat=1.9}
\def\BibTeX{{\rm B\kern-.05em{\sc i\kern-.025em b}\kern-.08em
    T\kern-.1667em\lower.7ex\hbox{E}\kern-.125emX}}
\begin{document}

\title{Towards Mobile Distributed Ledgers}

\author{
Dimitris Chatzopoulos, Anurag Jain, Sujit Gujar, Boi Faltings, and Pan Hui
\IEEEcompsocitemizethanks{\IEEEcompsocthanksitem Dimitris Chatzopoulos (dcab@cse.ust.hk) is with the Department
of Computer Science and Engineering, The Hong Kong University of Science and Technology, Hong Kong. 
\IEEEcompsocthanksitem Anurag Jain (anurag.jain@research.iiit.ac.in) and Sujit Gujar (sujit.gujar@iiit.ac.in) are with the International Institute of Information Technology, Hyderabad, India. 
\IEEEcompsocthanksitem Boi Faltings (boi.faltings@epfl.ch) is with the Ecole Polytechnique Federale de Lausanne, Switzerland.
\IEEEcompsocthanksitem Pan Hui (panhui@cse.ust.hk) is with with the Department of Computer Science and Engineering, The Hong Kong University of Science and Technology, Hong Kong and with the Department of Computer Science at the University of Helsinki, Finland. \protect }
\thanks{Manuscript received: date; revised: date}
}

\IEEEtitleabstractindextext{%
\begin{abstract}
Advances in mobile computing have paved the way for new types of distributed applications that can be executed solely by mobile devices on device-to-device (D2D) ecosystems (e.g., crowdsensing). Sophisticated applications, like cryptocurrencies, need distributed ledgers to function. Distributed ledgers, such as blockchains and directed acyclic graphs (DAGs), employ consensus protocols to add data in the form of blocks. However, such protocols are designed for resourceful devices that are interconnected via the Internet. Moreover, existing distributed ledgers are not deployable to D2D ecosystems since their storage needs are continuously increasing. In this work, we introduce and analyse Mneme, a DAG-based distributed ledger that can be maintained solely by mobile devices. Mneme utilizes two novel consensus protocols: Proof-of-Context (PoC) and Proof-of-Equivalence (PoE). PoC employs users' context to add data on Mneme. PoE is executed periodically to summarize data and produce equivalent blocks that require less storage. We analyze Mneme's security and justify the ability of PoC and PoE to guarantee the characteristics of distributed ledgers: persistence and liveness. Furthermore, we analyze potential attacks from malicious users and prove that the probability of a successful attack is inversely proportional to the square of the number of mobile users who maintain Mneme.
\end{abstract}

}

\maketitle

\begin{figure*}[t]
    \begin{subfigure}{\columnwidth}
        \centering
        \includegraphics[width=0.85\columnwidth]{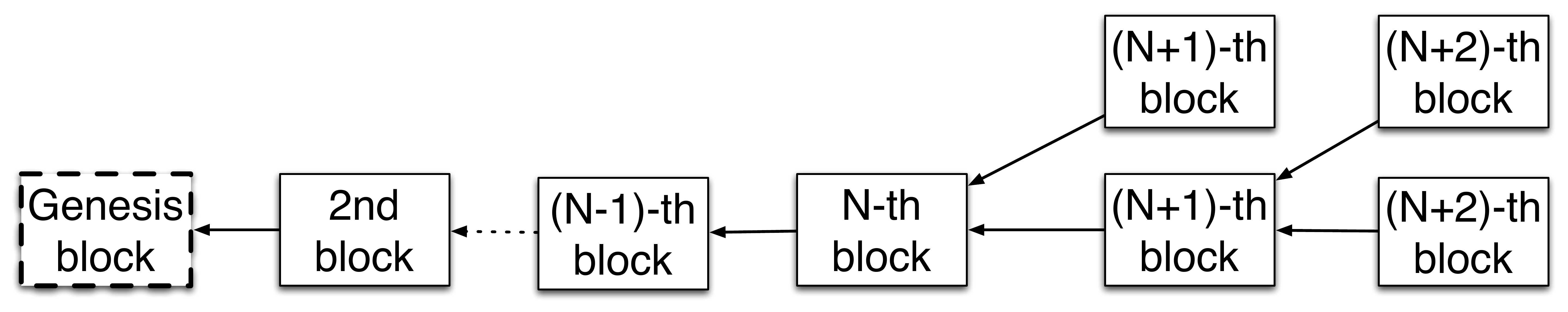}
        \caption{Blockchain.}
        \label{fig:fork}
    \end{subfigure}
    \begin{subfigure}{\columnwidth}
        \centering
        \includegraphics[width=0.85\columnwidth]{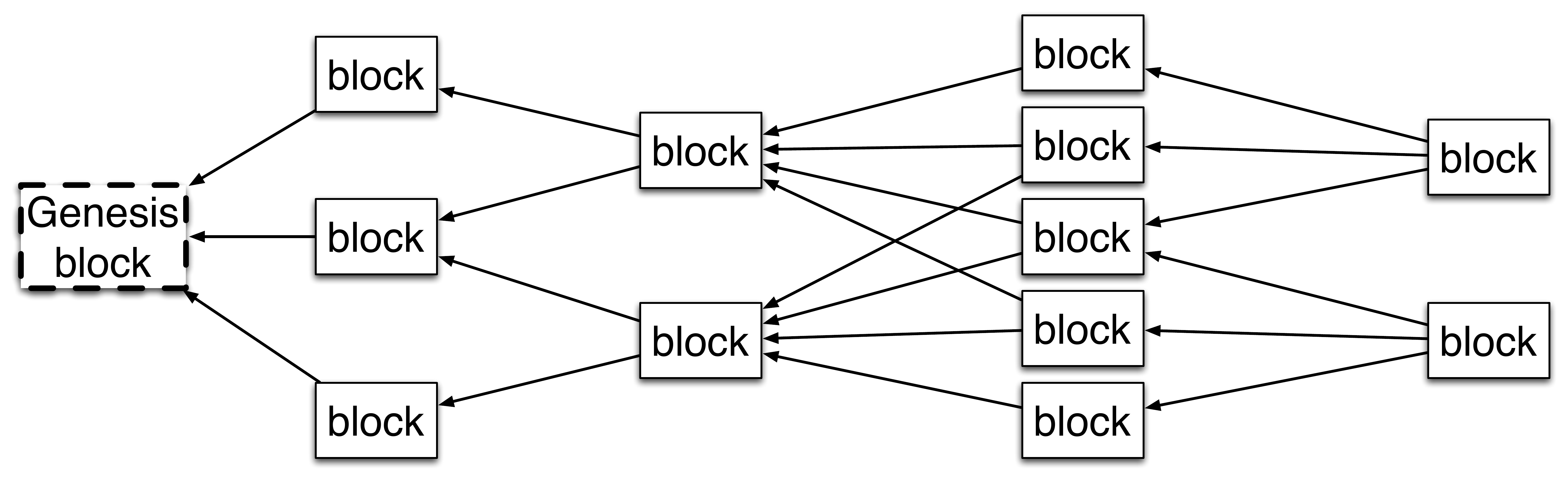}
        \caption{Directed Acyclic Graph.}
        \label{fig:dag}
    \end{subfigure}
    \caption{Representation of a blockchain with two forks (left) and of a DAG (right).}
\end{figure*}

\section{Introduction}
The popularity of permissionless distributed ledgers (DLs) increased with the development of Bitcoin in 2009~\cite{nakamoto2012bitcoin}. They store replicated data and are maintained by interconnected \emph{nodes} that exchange messages, which can join and leave the system at any time and are self-interested or even malicious. Although the most popular data type for DLs are transactions between users of cryptocurrencies, but they can also be used to store healthcare data \cite{fan2018medblock}, collected data from IoT devices \cite{popov2016tangle}, votes\cite{mccorry2017smart}, ownership titles and others. The two most popular types of DLs are the blockchain and the directed acyclic graph (DAG). DLs need to have two properties to be functional: \emph{persistence} and \emph{liveness}~\cite{cryptoeprint:2014:765}. 

Persistence measures how common is the view of the DL among the consensus nodes while liveness is associated the ability of the consensus nodes to add new data on the DL. Persistence is essential to ensure that credits are final and that they happened at a particular ``time'' in the system's timeline (implicitly defined by the DL itself). Liveness property ensures that the ledger makes progress by including new transactions. Consensus nodes eventually need to reach an agreement to ensure validity a DL. Depending on the design of a DL, they are rewarded for dedicating resources for its function. Bitcoin, Ethereum, and most cryptocurrencies, are using \textit{Proof-of-Work} (PoW) consensus protocol. Consensus nodes in PoW protocols (\emph{miners}) are rewarded proportionally to the processing power they contribute. In \textit{Proof-of-Stake} (PoS), another well-known protocol, consensus nodes (\emph{validators}) are rewarded based on the stake they own. Miners and validators store locally the DLs whose storage needs are increasing since data can only be appended.

\smallskip\noindent\textbf{Motivation.} Mobile devices can not operate as consensus nodes in existing distributed ledgers due to their \emph{(i)} poor processing capabilities compared to miners, \emph{(ii)} unstable connectivity, and \emph{(iii)} limited storage capacity. 
Motivated by the advancement in mobile devices capabilities, the plethora of network interfaces~\cite{gomez2012overview,camps2013device,mumtaz2014direct,mark2015peer}, the energy efficient link layer protocols, neighbor discovery protocols and broadcasting on infrastructure-less networks~\cite{asadi2013wifi,shen2016secure,funai2017enabling}, we argue that a distributed ledger can be maintained solely by mobile devices if its function is guaranteed by protocols that are based on the features of mobile devices.

\smallskip\noindent\textbf{Contributions.} We introduce Mneme, a DAG-based DL that is maintained solely by mobile devices, called \emph{corroborators}, and is based on two consensus protocols that utilise the characteristics of mobile devices: \emph{Proof-of-Context} (PoC) and \emph{Proof-of-Equivalence} (PoE). PoC is using users' context and reputation in its function to add blocks in Mneme. PoE detects blocks that can be deleted from Mneme and produces \emph{regenesis blocks}. Both protocols are based on unforgeable proofs and terminate when a big fraction of the corroborators is informed. 
Analysis using random geometric graphs and cryptographic protocols show that the persistence and liveness properties are guaranteed in Mneme.

\smallskip\noindent\textbf{Applications of Mneme.}
The first application of Mneme is as a cryptocurrency in ``underbanked'' areas where people do not have financial footprint and Internet connectivity is limited. Advances on opportunistic computing highlight the need for automated payment schemes between mobile devices that assist each other in the execution of tasks and need to be compensated~\cite{8024034,7509388}. The reduced connectivity on remote and rural areas has motivated the development of digital payment applications between mobile users but none of the existing proposals focuses on the security of the ledger that stores the transactions. For example, the authors of \cite{2018arXiv180110295H} propose a blockchain-based payment scheme in a similar way to Mneme, but they assume intermittent connectivity to a bank. Their proposal is based on a system that is composed of smart contracts while their blockchain is not maintained by mobile users. The authors of~\cite{2017arXiv170808086C} proposed a payment scheme that functions only with mobile devices, but they do not analyze the security properties of the underlying distributed ledger. 
Solutions similar to Mneme are already in use in mobile cloud computing architectures~\cite{pan2018edgechain,8024034}. In crowdsourcing, Mneme can be used to store the data generated by the mobile devices together with the additional information of the participants~\cite{farm}. Furthermore, a popular technique to increase the anonymity of the users on cryptocurrencies is a laundry service~\cite{miers2013zerocoin}.  

\section{Background}
\label{sec:background}
The breakthrough of Bitcoin is the use of the blockchain and PoW as a solution to the Byzantine Generals problem~\cite{lamport1982byzantine}, a classic problem in the distributed consensus literature. 
Formally, the problem of distributed consensus is: ``Given a number of nodes and assuming that each node has an input value and a subset of the nodes may be malicious, a distributed consensus protocol has two properties: \emph{termination} and \emph{agreement}''.
The first property implies that the protocol terminates with all the honest nodes agreeing on the same value and the second that the value the nodes have been agreed on is generated by an honest node. 
Satoshi Nakamoto, via PoW, violated the traditional assumptions upon which the impossibility results were built~\cite{fischer1985impossibility} by introducing incentives and randomization. The two most commonly employed protocols on permissionless DLs are PoW and PoS and are designed for nodes that can exchange messages via the Internet. In both of them, an election determines the next node to propose a block in the blockchain. 
In PoW, the election is based on solving cryptographic inequalities~\cite{dwork1992pricing}, while in PoS it is based on a cryptographic protocol that creates randomness~\cite{kiayias2017ouroboros}. 

PoC and PoE are designed for ecosystems where mobile devices exchange messages opportunistically forming a partially synchronous system. Their termination depends on the topology of the ecosystems and the introduced incentives. Their agreement is based on the created proofs. Consensus protocols designed for mobile ad-hoc networks cannot be utilised to maintain a distributed ledger because \textit{(i)} they assume that every node is aware of the considered inputs and \textit{(ii)} are based on message exchange~\cite{moniz2010turquois,brzezinski2012providing}. These make the performance of such protocols very poor since they require many messages to reach~consensus.

\smallskip\noindent\textbf{Distributed Ledgers.} The design of every distributed ledger has to guarantee persistence and liveness via two consensus protocols. The first protocol adds data to the ledger and the second verifies and allows their use. The blockchain is the most popular distributed ledger and is composed of blocks of data stored in sequence. A less popular type is directed acyclic graph (DAG), where the nodes of the graph can be transactions~\cite{cryptoeprint:2016:871,churyumov2016byteball,lerner2015dagcoin,popov2016tangle} or blocks~\cite{cryptoeprint:2016:1159,cryptoeprint:2018:104}. DAGs are employed to increase the processing capacity of the ledger and are part of the so-called second-generation of distributed ledgers. Both ledgers require a genesis block, which is the first block in the ledger, as a point of reference for what is a correct ledger. Depending on the employed consensus protocol, the genesis block may contain additional information. Ouroboros PoS protocol, for example, considers a genesis block that contains the public-keys of the stakeholders, their respective stakes and auxiliary information that is used to seed a leader election process~\cite{kiayias2017ouroboros}.

\smallskip\noindent\textbf{Blockchain.} PoW and PoS require an honest majority to be functional. The property of persistence is guaranteed via the longest chain rule. In PoW it is probable for more than one miners to produce a block that extends the same block and create a fork. Forks are caused by the absence of coordination between the miners. Figure~\ref{fig:fork} depicts three blockchains with two forks. The property of liveness is guaranteed by the fact that malicious nodes can not prevent normal users from generating blocks. A block is verified if it is in a chain with at least six consecutive blocks. This rule prevents race and Finney attacks~\cite{finneyfinney}. The security of a blockchain that uses PoW relies on honest nodes being sufficiently connected so that when one miner extends the chain with a new block, it propagates it in time to all honest nodes before the next one is created. To guarantee this property, the creation of blocks is regulated via the difficulty parameter in PoW. In Bitcoin, for example, a block is created every 10 minutes, on average. The low scalability of blockchain is the primary motivating factor for using DAG-based distributed ledgers.

\smallskip\noindent\textbf{Directed Acyclic Graphs.} DAGs store all the generated blocks (or transactions) in the graph, but they employ more sophisticated protocols of higher complexity to verify them~\cite{cryptoeprint:2018:104}. Each node of the DAG, as depicted in Figure~\ref{fig:dag}, is connected to at least two other nodes and each directed link from one node to another implies that the former verifies the later. Invalid blocks can be added on the DAG, but the other nodes will not verify them. Persistence and liveness are guaranteed by the voting between the DAG nodes that depends on how well each block is connected to the others.
 
\begin{figure}[t]
    \centering
    \includegraphics[width=0.85\columnwidth]{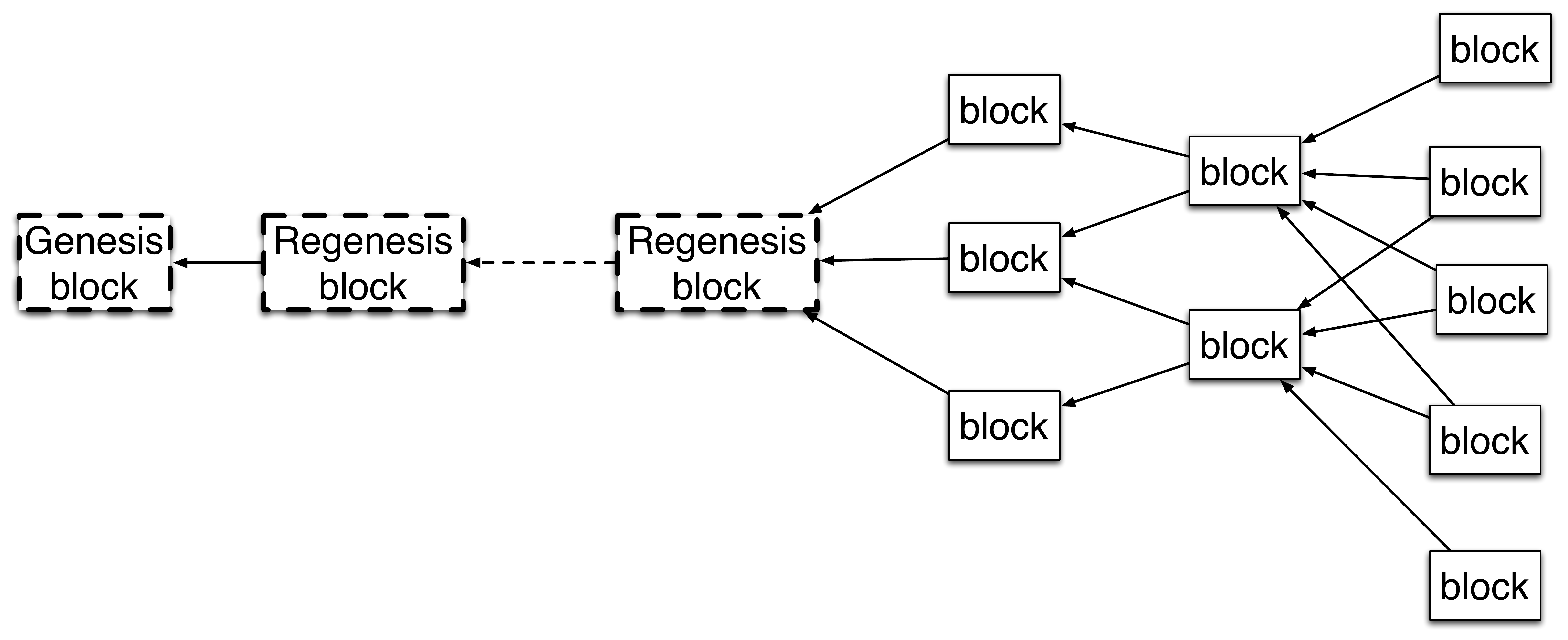}
    \caption{Structure of Mneme.}
    \label{fig:hydag}
\end{figure}

The main weakness of blockchains is the need for the corroborators to have the same view of the ledger. Given that we focus on an ecosystem where mobile device exchange messages opportunistically, this requirement slows-down the addition of data to the ledger. Mneme, as depicted in Figure~\ref{fig:hydag}, has the form of a particular type of DAG that is composed of two types of blocks. The conventional blocks that are created via PoC and the blocks that are generated by a subset of users via PoE. We call them \textit{regenesis blocks}, and use them as synchronization points. Regenesis blocks trigger the deletion of conventional blocks for releasing storage space. The blocks between regenesis blocks are deleted, and the included transactions are reorganized in new blocks hash pointers to the last regenesis block.

\section{Overview}\label{sec:design}

We design Mneme to store blocks of transactions between mobile users and data produced by them without the assistance from remote servers. Mobile users, named \emph{corroborators}, share their resources for the maintenance of Mneme to collect fees that are either generated by Mneme, via the PoE protocol, or are exchanged by mobile users who want to add data to the ledger. We introduce four types of fees that can be collected by the corroborators: \textit{transaction fees}, \textit{block creation fees}, \textit{block deletion fees} and \textit{block forwarding fees}. 

Every distributed ledger that stores transactions of credit, need to guarantee that malicious users can not perform \emph{double-spending attacks} while all the users can confirm the ownership of their credit. The defence against double-spending attacks is based on the consensus protocols that are responsible for the addition of new blocks in the ledger. The confirmation of credit ownership is demonstrated via hashpointers to verified blocks in the ledger.  
Mneme's operation is based on PoC and PoE consensus protocols. PoC forces corroborators to produce proofs of their context before processing and sharing blocks with their neighbors. PoC terminates when a block has been accepted by enough mobile users that are scattered in the area of deployment. The idea behind the design of PoC is to not accept a block as verified before a big fraction of the corroborators is aware of its existence. PoC is designed in such a way to satisfy persistence and liveness. The parameters of PoC determine the rate with which blocks are added to Mneme and the difficulty a malicious user will face in adding a conflicting block. PoE is executed periodically and produces regenesis blocks that can be used by users to synchronize with others and delete blocks from Mneme to released storage. PoE is based on \textit{proofs of equivalence} that are composed of two sets of blocks whose transactions produce equivalent balances to the involved accounts. PoE runs periodically to produce the next regenesis block. A subset of the corroborators is randomly selected, based on the fees they have collected, to execute the protocol. The design of Mneme is not based on any assumption regarding the capabilities of the devices. We only assume that mobile users are browsing within a predetermined geographic area and that there is an upper bound in the time needed for a message to be transferred between any two devices.

\subsection{System Model}
We consider a set of $\mathcal{N}$ corroborators who are browsing in an area $\mathcal{A}$. Each corroborator, $i \in \mathcal{N}$, has a public/private key pair ($pk_i$,$sk_i$) and is identified by her public key. Mobile devices can transact with each other to transfer credit and add data to Mneme. Each transaction contains fees that will be allocated to the corroborators who assisted on the process. Non-financial applications, like crowdsourcing, can use Mneme to exchange data between mobile users~\cite{mass}. In such scenarios, the collected credit by the corroborators should be exchangeable by the application. 

The corroborators maintain a distributed ledger $\mathcal{D}$ and, at time $t$, each one of them has her own view $\mathcal{D}_{i}(t)$. We expect corroborators to join and exit $\mathcal{A}$ arbitrarily. A mobile user $i$ is considered \textit{active} and can participate in the maintenance of  $\mathcal{D}$ whenever she is in $\mathcal{A}$, otherwise the user is \textit{inactive}. Active users at time $t$ are denoted by $\mathcal{N}_{a}(t)$. Inactive users that become active can get synchronized by requesting the last regenesis block and collecting accepted transactions that will be added to new blocks. For each user $i$, we denote her context by $c_{i}(t)$ and her reputation by $r_{i}(t)$, at time $t$. The context of a mobile user depends on her surroundings and can be measured via the sensors of her mobile device. The location of a mobile user $i$ is part of her context and is denoted by $c_{i}^{l}(t)$. Similarly, her neighbors are denoted by $c_{i}^{\mathcal{N}}(t)$. Two users are neighbors if the distance between them is less than a communication threshold to allow a message exchange.  The reputation of a user is calculated periodically and added to the regenesis blocks. We measure users' reputation based on the fees that they managed to collect since they are a clear measure of how much they have helped others. Trusted beacon devices and fingerprinting methods can also be employed to offer this functionality~\cite{xu2016device}. PoC and PoE are designed in such a way to tolerate a subset of the users to be \textit{malicious} (also known as Byzantine users). We denote the malicious users by $\mathcal{M}$ and their fraction by $f = |\mathcal{M}|/|\mathcal{N}|$.

\subsection{Network Model}\label{sec:NetModel}
As an underlying network that allows corroborators to exchange messages, we consider a mobile ad hoc network~\cite{macker1999mobile}, where mobile users are communicating opportunistically in a device-to-device manner~\cite{8594700}. 
We assume a partially synchronous network where at any time different users can have a different view of the ledger but with a common subset that includes the oldest blocks. 
Formally, there exist a timestamp $t_{sync} < t$ such that:
\begin{equation*}
    \mathcal{D}_{i}(t_{sync}) = \mathcal{D}_{j}(t_{sync}) \text{, }  \forall i,j \in \mathcal{N}, \{c_{i}^{l}(t), c_{j}^{l}(t)\} \in \mathcal{A},  i \neq j, 
\end{equation*}
$t_{sync}$ can be calculated via mobility prediction models~\cite{hsu2007modeling,su2004user,su2001mobility}. It is worth mentioning that the usual assumption on network synchronicity is to assume that the network is synchronous (i.e., if an honest user broadcasts a message, then all honest validators receive the message within a known maximum delay)~\cite{cryptoeprint:2017:406,kiayias2017ouroboros,Luu:2016:SSP:2976749.2978389,nakamoto2012bitcoin}. 
We can not make this assumption in the examined setting because mobile devices, depending on the size of $\mathcal{A}$, become inactive more frequently than miners and validators in conventional Internet-connected networks that maintain distributed ledgers. Also, depending on the fraction of the malicious users and their actions a mobile user may fail to receive a transaction. We assume that every message is delivered to every active user in less than $\Delta$. Any user who has not receive a message within $\Delta$ of its broadcast is considered inactive. We do not consider any fixed infrastructure, but the proposed protocols are readily adaptable to areas with such infrastructure. 


\subsection{Transactional Model}
Mobile users can create transactions and broadcast them to their neighbors who will forward them until they reach the recipient. Every transaction, apart from the public keys of the sender and the receiver and the exchanged amount, has as input hash pointers to the blocks that the sender wants to use to justify the ownership of her funds. Honest users are motivated to forward transactions because they collect transaction fees. However, a transaction cannot be added to the blockchain unless it is acknowledged by the receiver. This induces a meaningful delay in the acceptance of the transaction that allows the entire network to receive the transaction and detect any possible double spends happening simultaneously.

\label{transaction_acceptance_rule}
The receiver of one transaction will accept the transaction if she receives the transaction signed by at least \textit{a minimum number of trusted users} of her trusted network, denoted by $\mathit{mTr}$ and atleast $\delta$ time has elapsed since she had first received the transaction. Where $\delta$ can be chosen by the user that is going to accept the transaction. Letting the users pick $\delta$ introduces flexibility since the user may decide to accept a transaction quickly if it is of relatively low value or perceived as less risky (coming from a trusted user). \footnote{$\delta = \Delta$ assures complete safety against a double spending attack.}

The authors of~\mbox{\cite{2017arXiv170808086C}} show that such an acceptance rule is sufficient to prevent simultaneous double spending attacks as long as there are no disjoint components in the network. In our analysis, we show that the probability of such an event is negligible. Therefore, once the receiver acknowledges the transaction, he/she can be confident that no honest user has accepted a conflicting transaction.

\par Each user stores acknowledged transactions locally in her pool of pending transactions $\mathcal{P}_{i}(t)$. Any user can propose a new block $b$ by adding $B$ transactions in it and triggering PoC. PoC is described in detail in Section~\ref{sec:poc}. PoE, on the other hand, is executed periodically every $T$ time units (epoch), by a randomly selected set of users $\mathcal{K}_{\tau} \subset \mathcal{N}_{a}(\tau\cdot T)$. Where $\tau$ denotes the epoch index. The probability of a user to be selected in each epoch is proportional to her reputation. PoE terminates successfully if a minimum number of users, $\mathcal{K}_{\tau}^{m}$ reach an agreement.
To demotivate malicious users to attack Mneme, we define a special type of transaction named \textit{conditional self transaction}. These transactions are based on Proof-of-Burn policies~\cite{proofofburn}. Every user who participates either in PoC or PoE creates a conditional self-transaction with a number of credits that she will lose if she behaves maliciously. 


\subsection{Incentives}\label{sec:incentives}
Incentives are required to balance the energy loss through the participation in the maintenance of the Mneme~\cite{banerjee2007energy}. Users should be allowed to collaborate (e.g., like the miners of the Bitcoin mining pools), while their incentives are aligned with the protocols~\cite{lewenberg2015bitcoin}. Although the focus of this manuscript is the security guarantees of Mneme and not the engineering of the incentives required to guaranty Mneme's function, we use this paragraph to discuss the incentives needed without further analysing them. We list four types of incentives: transaction fees, block creation fees, block deletion fees and block forwarding fees. We introduce them motivated by the extensive literature on incentives in opportunistic networks and by their importance on the design of cryptocurrencies. 

\smallskip\noindent\textbf{Transaction fees.} Users who produce transactions add fees that will be collected by others. Similarly to conventional cryptocurrencies, the higher the fees, the faster a transaction is expected to be added to Mnene~\cite{PredbitcoinFees}. Every transaction is signed by users who forwarded it before its broadcast. 

\smallskip\noindent\textbf{Block forwarding fees.} Users who assist on the block creation or deletion processes share a fraction $\phi_c$ of the block creation fees and a fraction $\phi_d$ of the block deletion fees. 

\smallskip\noindent\textbf{Block creation fees.} To motivate users to store locally accepted transactions and add them to blocks, the user who produces it adds a block fee that will be collected by other users. A fraction $1-\phi_c$ of block fees of the transactions that are included in a produced block are shared among the users who produced the block. 

\smallskip\noindent\textbf{Block deletion fees.} To motivate users to increase their reputation, every regenesis block adds new credits to the system and assigns them to the users who run the PoE. Then these selected users produce a new block to share a fraction $\phi_d$ of their earnings with the users who forwarded the block deletion messages. The probability of a user to be selected to run PoE is proportional to the fees she has collected.

\subsection{Proofs-of-Context}\label{sec:proof1}
The functionality of Mneme depends on the users' context and primarily on their location that needs to be robustly estimated and verified by neighboring devices. To prove a user $i$ that her location is as measured by her GPS or any localization method~\cite{bulusu2000gps,Capkun2002}, she uses the cryptographic protocol presented below:

\smallskip\noindent\textbf{1) Scanning.} Scan for neighbours and produce a location message to interact with them: $m_{i}(t) = \left\{c_{i}^{l}(t),c_{i}^{\mathcal{N}}(t) \right\} $

\smallskip\noindent\textbf{2) Tag production.} Use $m_{i}(t)$ to produce a tag of fixed size, $tag_{i} = f\left( m_{i}(t)\right)$, via a pseudo-random function~\cite{impagliazzo1989pseudo} stored in the genesis block. A popular example is HMAC~\cite{krawczyk1997hmac} that is used in SSL, SSH, etc. that produces tags of 256 bits. 

\smallskip\noindent\textbf{3) Commitment.} Use the secret key $sk_i$ to produce a commitment for every neighbor $\Upsilon_{i}(t)$: \vspace{-0.1cm}
\begin{equation}
	\mathcal{C}omm \left(m_{i}(t), tag_{i}\right)  \xlongrightarrow{sk_i} \Upsilon_{i}(t),
\end{equation}

\smallskip\noindent\textbf{4) Validation.} Every neighbour receives $\Upsilon_{ij}(t)$ and examines whether user $i$ is at $c_{i}^{l}$ at time $t$. \vspace{-0.1cm}
\begin{equation}
	\mathcal{V}er\left(m_{i}(t), tag_{i}, \Upsilon_{ij}(t)\right) \xlongrightarrow{sk_j} \Omega_{ji}(t) \in \{\texttt{yes}, \texttt{no}\}.
\end{equation}
$\Omega_{ji}(t)$ equals to ``\texttt{yes}'' if user $j$ verifies that user $i$ is her neighbour (i.e., their locations differ by less than a threshold) and ``\texttt{no}'' otherwise. Every user, after receiving $\Upsilon_{i}(t)$ can use the public key of $i$ to extract the location user $i$ claims to be at time $t$ and her neighbours together with $tag_i$. User $i$, by sending $\Upsilon_{i}(t)$ instead of $m_{i}(t)$ makes sure that her neighbors can only answer to her claim. Any malicious user is not able to change the location user $i$ claims to be. The integrity of $\Upsilon_{i}(t)$ is guaranteed by the use of a pseudo-random function in the production of the the tag~\cite{stubblebine1992message}. Practically, a malicious user can only produce a PoC for a location she is not currently in. By doing that, she will not be able to verify her fake location by normal users. Malicious users can still assist each other, and for that reason, we add a reputation weight to PoCs that depends on the reputation the neighbors. 

Via this process user $i$ can construct a \textit{Proof-of-Context} that a set of her neighbors argues that are within a given distance threshold from her at time $t$. 
\begin{equation}
    \Pi_{i}^{C}\left(\Upsilon_{i}(t), \bigcup_{j \in c_{i}^{\mathcal{N}}(t)} \Omega_{ji}(t),  \sum_{j=1}^{|c_{i}^{\mathcal{N}}(t)|} \frac{1_{\{ \Omega_{ji}(t)==``\texttt{yes}''\}}r_{j}(t)}{|c_{i}^{\mathcal{N}}(t)|}\right)
\end{equation}

PoC is defined as the set of messages from the neighbouring devices of a user that the user is at a specific location. Each message is signed by the neighbouring users, and their worth is associated with their reputation.

\subsection{Proofs-of-Equivalence}\label{sec:proof2}

Every $T$ time units PoE is triggered to create a regenesis block and at the $\tau$-th round users in $\mathcal{K}_{\tau}$ are randomly selected to produce the block. Each of the selected users $i$ is responsible for collecting all the blocks that are added in the ledger during the $\tau$-th round, $\mathcal{D}_{\tau}$ and create a smaller number of blocks with equivalent outcome $\mathcal{L}_{\tau}^{i}$ and produces a proof-of-equivalence:
\begin{equation}
    \Pi_{i}^{E}\left(\mathcal{D}_{\tau},\mathcal{L}_{\tau}^{i}\right)
\end{equation}

\noindent\textbf{Example.} We consider two blocks with four transactions each between $Alice$, $Bob$, $Carol$ and $David$. We assume that $Alice$, $Bob$ and $David$ are selected to summarise the two blocks and that they will share three credits as block fees. The eight transactions are:
$tr(Alice\to Bob) = 5$,  $tr(Alice\to Carol) = 2$, $tr(Alice\to David) = 2$,  $tr(Bob\to David) = 1$, $tr(David\to Carol) = 2$, $tr(Bob\to Alice) = 1$, $tr(Carol\to Alice) = 1$ and$tr(Carol\to David) = 1$. The balance changes are $Alice\searrow7$, $Bob\nearrow3$, $Carol\nearrow2$, $David\nearrow2$, where the upright pointing arrow denotes an increase and the downright a decrease. We can now create one block and store four transactions where Alice will transfer to a \textit{virtual user} $7-1$ coins and the virtual user will transfer $3+1$ coins to Bob, $2$ coins to Carol and $2+1$ coins to David. Via this process we can delete the two blocks that contained the eight transactions and add one with four. 

Alice, Bob, and David, will generate this block in the same way as a conventional blocks using PoC and add it to the regenesis block. The only difference is that only the selected users are allowed to sign and verify it. Each of them is allowed to assign newly mined credits to herself. If the users do not have the same view, some of the blocks will not be created, and their creators will lose their credits. This motivates them to make sure that they are aware of every block needed before the initiation of the process. The selected users should be more than the users needed to sign and verify one block. 

\section{PoC Consensus Protocol}\label{sec:poc}
Blocks are generated by mobile users and contain $B$ transactions each. Every corroborator $i$ can propose a new block $b$, at time $t$, by using $B$ of the transactions she has witnessed and stored locally in $\mathcal{P}_{i}(t)$. We denote the transactions that are added to $b$ by $\textbf{tr}$, $|\textbf{tr}| = B$ and the $j$-th transaction in $\textbf{tr}$ by $\textbf{tr}_{j}$. The broadcast of one block requires $\mathcal{O}(\mathcal{N})$ messages if the topology is changing frequently~\cite{newport2017gossip}. Each block, after its creation, is connected to a number of existing blocks in Mneme via a set of hash pointers. These hash pointers are determined by the user $i$ who initiated the creation of the block and are the hashes of the blocks, in $i$'s view, that does not have any incoming link in $\mathcal{D}_{i}(t)$. Each block after being verified by a set of corroborators, is added to Mneme. The users are sharing the same source of randomness that is determined in the genesis block and is updated on every regenesis block~\cite{cryptoeprint:2017:216},~\cite{kiayias2017ouroboros}.

Honest users are expected to receive a new block within a given period, $\Delta$. This means that the difference between the creation times of two contradictory blocks should be less or equal to a given period. If this difference is higher, there will be a path between the two blocks unless the second block was created by a malicious user who intentionally selects to not connect it to blocks that have a path to first one. Using a verification algorithm similar to the one proposed by~\cite{cryptoeprint:2016:1159} we can decide which of the two blocks to keep. The block creation process is composed of three phases and can be triggered by many mobile users at any time: 

\smallskip\noindent\textbf{$P_{0}^{PoC}$\big]} \textbf{Block Creation and Broadcasting.} A user creates a block using $B$ transactions and broadcasts it. Each block, apart from transactions and hashes to existing blocks, has a variable that stores the average distance between the users that will sign block $b$, $\bar{l}_b$ and a set of user-context pairs: 
\begin{equation}
       \mathcal{D}_{b} = \{<i,\Pi_{i}^{C}>,<j,\Pi_{j}^{C}>,\ldots  \}. 
\end{equation}
At this phase, $\mathcal{D}_{b}$ contains the public key of the user who created the block and her PoC, while $\bar{l}_b$ is initialized to~0. 

\smallskip\noindent\textbf{$P_{1}^{PoC}$\big]} \textbf{Block Signing and Forwarding.} Every corroborator that receives a block with a valid PoC, examines the included transactions and if she is familiar with every one of them she signs and adds her PoC to the set of user-context pairs and forwards the block.

\smallskip\noindent\textbf{$P_{2}^{PoC}$\big]} \textbf{Block addition to Mneme.} A block is considered verified if at least $mRS$ users have signed it and the average distance between them is higher than $mD$. Whenever a user is the $mRS$-th user who receives a block, she examines whether the average distance between the users that have signed it is higher than $mD$. If this is true, she can use the locations of the users to construct a proof that verifies the block. If the users that have signed the block are not spread enough, the user rebroadcasts the block.  This requirement ensures that each block is spread across the network.  After the verification of the block, the users that have verified the block share a fraction of the block creation fees.
\par Any user that receives a verified block will accept the block if and only if none of the new transactions conflict with the set of transactions it had accepted earlier.
Malicious users may try to sign a block using a fake PoC to speed up the block creation process or try to cooperate with each other to produce blocks that violate the proof-of-ownership rule. The required PoCs do not allow users to spoof their location while any block that violates the proof-of-ownership rule will be discarded and the malicious users who produced it will only waste their resources and credits from their conditional self-transactions. 

Based on the analysis of SPECTRE~\cite{cryptoeprint:2016:1159} and by replacing PoW by PoS~\cite{kiayias2017ouroboros}, persistence and liveliness are still guaranteed. Note that, we also have extra security measures by incorporating Proof-of-Context, that is, every verifier of the block attaches its proof of location signed by her highly reputed neighbors. PoC scales with the number of the mobile users since the more the corroborators, the faster the messages are broadcasted, and the faster the blocks are created.

\section{PoE Consensus Protocol}\label{sec:del}

\begin{figure}[t]
        \centering
        \includegraphics[width=1\columnwidth]{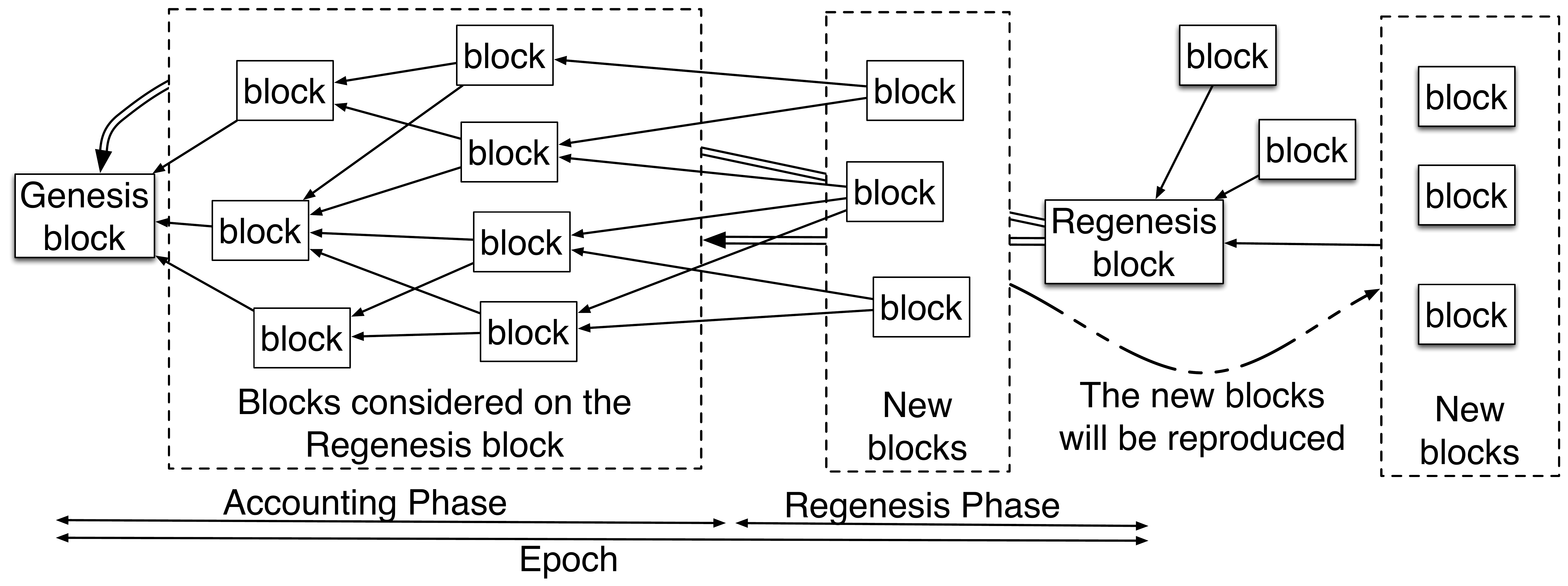}
        \caption{Schematics of PoE consensus protocol.\vspace{-0.3cm}}
    \label{fig:Regenesis}
\end{figure}

To reduce the storage overhead of Mneme we propose a solution to delete blocks and create \textit{regenesis blocks} that summarise the transactions and the data of the deleted ones. Regenesis blocks are created in epochs by a randomly selected subset of users. The probability of one user to be selected is proportional to her reputation. That is, after every $T$, the protocol describe below is executed in order to produce a new regenesis block. In each round $\tau$, the protocol selects $\mathcal{K}_{\tau}$ users out of $\mathcal{N}$ users with high reputation to validate the regenesis block, and the regenesis block is accepted only if at least $\mathcal{K}_{\tau}^{m}$ of the $\mathcal{K}_{\tau}$ agree to it. These users will also share a fraction of the new coins $\Xi_{\tau}$ that will be produced via this process. In case of not reaching a consensus within the epoch, the regenesis block is not created and the selected users of the next epoch are responsible to produce a regenesis block that summarizes all the blocks of the last two epochs. 
The regenesis block creation process is triggered periodically and the users that will be responsible are selected at the beginning of each epoch. The source of shared randomness and the reputation of each user allows the participating users figure out if they are selected or not. Figure~\ref{fig:Regenesis} depicts the whole process. 

\smallskip\noindent\textbf{Accounting Phase} $P_{0}^{PoE}$\big] Every selected user $i \in \mathcal{K}_{\tau}$ at round $\tau$ actively monitors for blocks $\mathcal{D}_{\tau}$ that are added in Mneme, detect all the involved accounts, summarise their balance change and produces $\mathcal{L}_{\tau}^{i}$. Via protocols like Spectre~\cite{cryptoeprint:2016:1159} each user is able to detect double spending attacks on the DAG. In parallel, the users calculate the fees each user collected in order to update the reputation scores. 

\smallskip\noindent\textbf{Regenesis Phase }$P_{1}^{PoE}$\big] A regenesis block is created and is verified by the selected users. The reputation of each user is updated based on the fees that were allocated to her in the DAG. The regenesis block has a pointer to the previous regenesis block and to the header of every block in the DAG that was considered by the protocol. After its verification, the regenesis block is broadcasted and every user that receives it is able to delete the blocks that has stored locally and are included on the regenesis block. Depending on the number of the addresses involved on the examined DAG and the number of the transactions that can fit in a block, a number of blocks are created and are verified via PoC consensus protocol in the same way as being new blocks. In order to minimize the generated traffic, only a subset of the selected users are able to initiate this process and verify the new blocks. 

\begin{table}[t]
\small
    \begin{center}
    \begin{tabular}{ p{1cm}  p{7cm} }
    \toprule
    \textbf{Symbol}      &  \textbf{Description}    \\
      \cmidrule(r){1-2}
      $B$              & Number of transactions per block\\
      $mRS$        & Minimum number of signatures that are required for a block to be added to the ledger.  \\
      $mD$          & Minimum average distance between the corroborators that sign a block creation. \\
      $T$         	& Time between two regenesis blocks. \\
      $\delta$      & Time receiver waits before accepting a transaction (determined by an individual user). \\
      $\Delta$      & Maximum delay for the transfer of a block to all honest nodes (determined by the environment).\\ 
       \bottomrule 
      \end{tabular}
    \caption{Relevant Notation\vspace{-0.5cm}}
    \label{tab:params}
    \end{center}
\end{table}

\section{Analysis}\label{sec:analysis}
To show that Mneme is functional, it suffices to show that it has two characteristic properties, which refer to the stored blocks: \textit{liveness} and \textit{persistence}~\cite{garay2017bitcoin}. These properties are guaranteed if the corroborators can add new blocks and if the produced blocks are broadcasted to every corroborator. Particularly, if PoC and PoE terminate and there exist an upper bound on the time a corroborator needs to get informed about a new event. In this section, we analyze the conditions under which PoC and PoE terminate, we introduced the considered threat model in the design of Mneme and we discuss Mneme's robustness against possible attacks. \\


\noindent\textbf{Termination of PoC.} Assuming that user $u$ triggered PoC to produce block $b$ at $b_{t}$, we prove that PoC terminates and $b$ is included in $\mathcal{D}$. 
We define $\rho_{b}^{u}$ as the probability of user $u \in \mathcal{N}_a(b_{t})$ to be able to sign block $b$: $\rho_{b}^{u} = \Pi_{j=1}^{B} Prob\left[\textbf{tr}_{j} \in \mathcal{P}_{u}(t)\right]$, where $Prob\left[\textbf{tr}_{j} \in \mathcal{P}_{u}(t)\right]$ is the probability of user $u$ to have stored transaction $\textbf{tr}_{j}$ in her pool of pending transactions. Under the assumption of a partially synchronous network, there exist a graph $\mathcal{G}(N,E)$ that shows how each user receives the block creation message. We assume that if one user received the same message by more than one users, she received it first by one who was able to sign it. This assumption is realistic if we force the users who are not able to sign a block creation message to back off for a few seconds before forwarding it. 
Each user is expected to have $\pi |\mathcal{N}_{a}(t)| \frac{R^{2}}{|\mathcal{A}|}$ neighbors~\cite{DBLP:journals/corr/abs-0706-1063}, where $R$ is the fraction between the coverage area of the technology used from users to communicate over the deployment area (i.e., if the users communicate via WiFi-direct, that have a coverage radius of 50 meters, in an area 500 meters by 500 meters, $R=0.1$). PoC terminates if the following problem has a feasible solution: 
\begin{eqnarray}
	 \min \limits_{\textbf{n}} && \frac{1}{{|\textbf{n}| \choose 2}} \sum \limits_{ \substack{u\neq v, \\ u,v \in \textbf{n}}} dist(u,v) \\ \vspace{-0.0cm}
\text{subject to: }	&& \frac{1}{{|\textbf{n}| \choose 2}} \sum \limits_{ \substack{u\neq v, \\ u,v \in \textbf{n}}} dist(u,v)  \geq  mD \notag \\ \vspace{-0.0cm}
	&& \sum\limits_{u=1}^{|\textbf{n}|} \sum\limits_{j=1}^{B} 1_{\textbf{tr}_{j} \in \mathcal{P}_{u}(t)} = |\textbf{n}|\cdot B, |\textbf{n}| \geq mRS \notag 
\end{eqnarray}
where $\textbf{n}$ is any subset of $\mathcal{N}_a(t)$ that can sign the block and are spread enough. The existence of the feasible solution depends on $\{\rho_{b}^{u}\}_{u \in \mathcal{N}_{a}(t)}$, $mRS$ and $mD$. $\rho_{b}^{u}$ depends on the incentives provided to the mobile users. High values of $mRS$ and $mD$ will delay PoC and cause the generation of many forwarding messages but they will provide higher security guarantees. 

\smallskip\noindent\textbf{Termination of PoE.} Assuming that $\mathcal{K}_{\tau}$ corroborators are selected to produce the new regenesis block. They need to collect all the created blocks of the epoch. We prove that PoE terminates and the new regenesis block is produced. We define $\theta_{\tau}^{u}$ as the probability of user $u \in \mathcal{K}_{\tau}$ to be able to produce the regenesis block of the period $\tau$: $\theta_{\tau}^{u} = \Pi_{j=1}^{D} Prob\left[b \in \mathcal{D}_{u}(\tau)\right]$, where $Prob\left[b \in \mathcal{D}_{u}(\tau)\right]$ is the probability of user $u$ to have stored block $b$ in her local pool of Mneme. At least $\mathcal{K}_{\tau}^{m}$ out of $\mathcal{K}_{\tau}$ need to agree on the proposal of the regenesis block and this will happen with probability:
\begin{equation}\label{eq:PoEterm}
P[PoE\_termination] = \frac{\mathcal{K}_{\tau}}{\mathcal{K}_{\tau}^{m}}(\theta_{\tau})^{\mathcal{K}_{\tau}}(1-\theta_{\tau})^{(\mathcal{K}_{\tau} - \mathcal{K}_{\tau}^{m})}
\end{equation}
assuming $\theta_{\tau}^{u} = \theta_{\tau}^{v}=\theta_{\tau}$, for any two corroborators $u$ and $v$. In the case of $\theta_{\tau}^{u} \neq \theta_{\tau}^{v}$, the number of the coroborators who will sign the new regenesis block can be approximated via a normal distribution with mean $\sum_{u \in \mathcal{K}_{\tau}}\theta_{\tau}^{u}$ and standard deviation $\sum_{u \in \mathcal{K}_{\tau}}\theta_{\tau}^{u}(1-\theta_{\tau}^{u})$. $P[PoE\_termination]$ can be then calculated using the cumulative distribution function of the approximated distribution and $\mathcal{K}_{\tau}^{m}$. 

\begin{figure*}[t]
\centering
    \begin{subfigure}{0.65\columnwidth}
    \centering
    \includegraphics{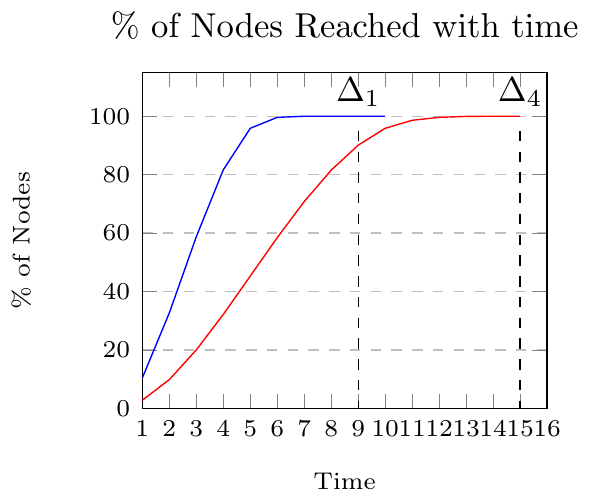}
      \caption{Spread of a message in a RGG for 1000 Users spread over 1 Sq. Km.}
      \label{fig:rgg_spread}
\end{subfigure}
    \begin{subfigure}{0.65\columnwidth}
    \centering
    \includegraphics{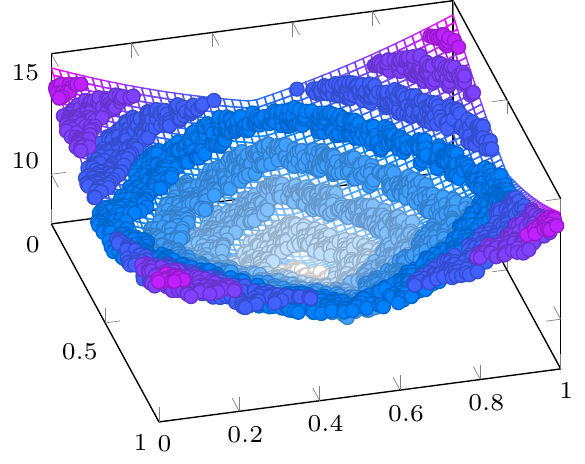}
      \caption{Variation of $\Delta$ with position of the user (in a setting with static users)}
      \label{fig:rgg_3d}
\end{subfigure}
    \begin{subfigure}{0.65\columnwidth}
    \centering
     \includegraphics{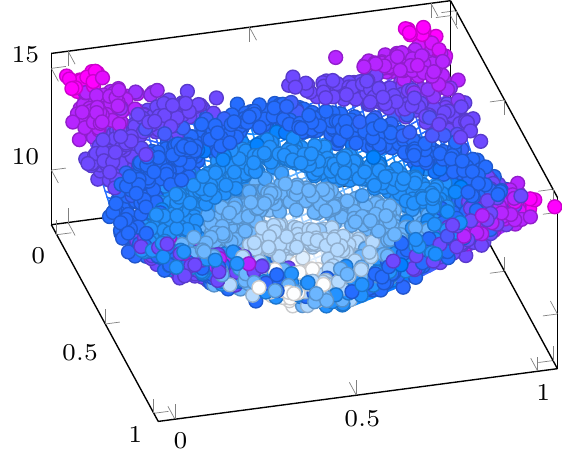}
      \caption{Variation of $\Delta$ with position of the user (in a setting with mobile users)}
      \label{fig:rgg_3d_2}
\end{subfigure}
    \caption{Variation of $\Delta$ along with parameters of Mneme and the position of a user\vspace{-0.2cm}}
        \label{fig:4}
\end{figure*}
\section{Threat Model}
\label{threat_model}
We denote the fraction of Byzantine corroborators by $f=|\mathcal{M}|/|\mathcal{N}|$ where $\mathcal{M} \geq mRS$ that may strategically deviate from the protocol. 
Their goal is to either steal credit or incapacitate Mneme. We expect them to act arbitrarily against PoC and PoE individually or in coordination. We assume that the resources of each of the Byzantine corroborators are equivalent to an honest corroborator. Practically, we expect Byzantine corroborators to: \mbox{\textit{(i)}} not forward messages, \mbox{\textit{(ii)}} generate fake transactions and blocks, \mbox{\textit{(iii)}} forge PoCs and PoEs, and \mbox{\textit{(iv)}} isolate honest corroborators. The most common attack is the double spending attack where malicious may assist each other to create two valid blocks with the same input transaction. By managing that, they will have successfully spent the same coins twice. Considering that Mneme allows the substitution of blocks with less blocks of an equivalent impact on the users' balances, Byzantine corroborators may try to claim ownership of coins that do not belong to them. Malicious users may collaborate to perform Eclipse attacks~\mbox{\cite{singh2006eclipse}} 
in order to isolate honest corroborators and spoof them to accept already spent transactions. Another type of a potential attack is the so-called Hijacking attack~\mbox{\cite{apostolaki2017hijacking}}, where a Byzantine corroborator creates fake messages to misinform honest corroborators a new block/transaction.

\subsection{Characteristics of Adversaries} 
To justify the robustness of Mneme, we consider adversaries with capabilities that normal participants do not have. We consider adversaries that have the following advantages over honest corroborators by deviating from the protocol.

\smallskip\noindent\textbf{Wormhole Attack.}
In a \emph{Wormhole Attack}, the attackers locate themselves strategically across the network and communicate across a secret channel. This might allow the adversary to transfer a message between geographically separated attackers without broadcasting the same to the network.

\smallskip\noindent\textbf{Unbounded Block Creation.}
Notice that since $\mathcal{M} \geq mRS$, the adversary can produce a valid PoC by signing a block with only its corroborators. The adversary could potentially distribute the corroborators geographically so that the average distance between them becomes greater than $mD$. The adversary could then collect both the private and public keys of the corroborators at a single location and produce the $mRS$ signatures required, without any network delays. Although, such an attack is not possible on other distributed ledgers like Bitcoin or SPECTRE, we show that this does not lead to a loss of security in Section \ref{double_spending_analysis}.
\subsection{Analysis of Possible Attacks}
\noindent\textbf{Attacking PoC via the Wormhole Attack.}
 There is a possibility that the corroborators are disconnected in disjoint components that are large enough to produce valid blocks. In such cases, a malicious user may try to double spend her coins, once in the first component and then in the second component. Using random geometric graph (RGG) theory, we can argue that such probability is inversely proportional to the square of the number of the corroborators. In more detail, given a 2-dimensional RGC that is composed of $\mathcal{N}_a$ uniformly distributed users that can exchange a message if the distance between them is at most $R$, the following holds~\cite{gupta2000capacity,penrose2003random}: For $\mathcal{N}_{a} R \geq 2\log \mathcal{N}_a$, there is a path between any two users with probability $1-\frac{1}{\mathcal{N}_{a}^2}$. Thus:  
\begin{equation}
P[Double\_spending] \leq	\frac{1}{\mathcal{N}_{a}^2}
\end{equation}
An active adversary may try to split a connected component by controlling devices in a particular region such that messages are not transmitted from one component to another. For this, the width of such region under control should be at least $2R$, and again under uniform distribution model, the fraction of nodes that active adversary should control is non-practical. The probability would reduces even further in a setting with mobile users that are actively moving around, with the possibility that a user that was part of one component moves to another component. Thus, if we can achieve a certain level of security with $n$ static users, we could achieve the same level of security with $m (<n)$ mobile users.\\
\subsection{Analysis of Double Spending Attacks} 
\label{double_spending_analysis}
\smallskip\noindent\textbf{Producing Acknowledgement for Double Spends.} 
We consider a double spending attack successful if an adversary manages to spend the same input in transactions to two honest nodes. In order to do this, the adversary would require $mTr$ users from both the nodes trusted set to sign the transaction without any of the recipients of the transactions receiving the other transaction. In case of a double spending attempt, the adversary would need to propose her transaction to $mTr$ users of another user before they receive the first transaction. If a trusted user detects the double spending attempt, she would alert the respective user who would then reject the transaction. We show that once a delay of $\delta$ has elapsed and $mTr$ trusted users have signed and forwarded the transaction, this transaction would have been sufficiently propagated throughout the network (Figure~\mbox{\ref{fig:rgg_spread}}). Thus, either the trusted user will register the transaction or detect a double spend. Therefore, the double spending attempt would be foiled by social engineering.

As observed in Figure~\mbox{\ref{fig:rgg_spread}}, by setting $\delta = 5$ times the time taken to transfer the message between two users, the user can ensure that 95\% of other honest users would receive the message. Hence, the probability of a double spending attack being successful would be negligible since it would not only require the other recipient to belong to the remaining 5\% but also atleast $mTr$ trusted user of that recipient to also belong to this set of users.
\smallskip\noindent\textbf{Faking an Acknowledgement.} 
Let us consider if in the previous scenario, one of the transaction is addressed towards the recipient and another recipient is the adversary itself. In this case, he/she would generate an acknowledgement for the transaction forcefully. In this case, even if the adversary manages to produce a block using PoC that contains the transaction addressed to herself, the block would be rejected by rest of the honest users that contain the original transaction in their $\mathcal{P}_{i}(t)$ or they had already accepted a block containing the original transaction.

\smallskip\noindent\textbf{Attacking PoE via Collusion:}
A malicious user may try to form collusion of $\mathcal{M}$ nodes to attack PoE. For a successful attack, at least $\lfloor{\frac{\mathcal{K}_{\tau}}{2}+1}\rfloor$ nodes should be selected from $\mathcal{M}$ to participate in the process while $\mathcal{K}_{\tau}^{m}$ are needed to steal credit. If the selected malicious corroborators are more than $\frac{\mathcal{K}_{\tau}}{2}$ but less than $\mathcal{K}_{\tau}^{m}$, they can prevent honest corroborators from building a new regenesis block. 
There are $|\mathcal{N}| \choose |\mathcal{K}_{\tau}|$ ways of nodes getting selected to validate a regenesis block. Out of which favorable to the adversary are: $\mathcal{M}\choose \lfloor{\frac{\mathcal{K}_{\tau}}{2}+1}\rfloor$ +$ M \choose \lfloor{\frac{\mathcal{K}_{\tau}}{2}+2}\rfloor$ + $\ldots$ + $ \mathcal{M} \choose \mathcal{M}$  instances. Since the total number of instances is less than $2^\mathcal{M}$, the probability of a successful attack is: 
\begin{equation}
 P[Credit\_stealing] \leq \frac{{\mathcal{N}_a\choose \mathcal{K}_{\tau}}}{ 2^\mathcal{M} } \approx \frac{\mathcal{K}_{\tau}! }{2^{M}\mathcal{N}_{a}^{\mathcal{K}_{\tau}}}
\end{equation}
This probability is high when $\mathcal{K}_{\tau}$ is small and $M$ is as large as possible. However, $P[Credit\_stealing]$ is decreasing dramatically by the use of reputation scores in the selection of the $\mathcal{K}_{\tau}$ corroborators. 

\smallskip\noindent\textbf{Example.} For $|\mathcal{N}_a|=100, |\mathcal{K}_{\tau}|=10$, and $ |\mathcal{M}|=10$,  the probability of successfully attacking PoE is $< 2^{-28}$. while for $|\mathcal{N}_a|=1000$, and $|\mathcal{K}_{\tau}|=\mathcal{M}=100$, it is  $< 2^{-475}$.

\smallskip\noindent\textbf{Impact of Mobility.}
The probability of a successful attack on PoC is decreasing dramatically by users' mobility since an active adversary will need to employ a higher fraction of the users in order to stop message forwarding. Additionally, Eclipse attacks, designed for peer-to-peer networks, are not feasible in Mneme since corroborators, in difference to traditional consensus nodes, change their connected peers frequently over time. Similarly, Hijacking attacks are infeasible since the corroborators are communicating via broadcasting and not unicast protocols like the ones used in Internet-based ledgers. We do not consider specific mobility patterns for the movement of the users in order to robustly secure Mneme and guarantee persistence and liveness even in cases where corroborators are relatively static or move arbitrarily.

\begin{figure*}[t]
\centering
    \begin{subfigure}{0.65\columnwidth}
        \centering
        \includegraphics[height=\columnwidth, angle=270]{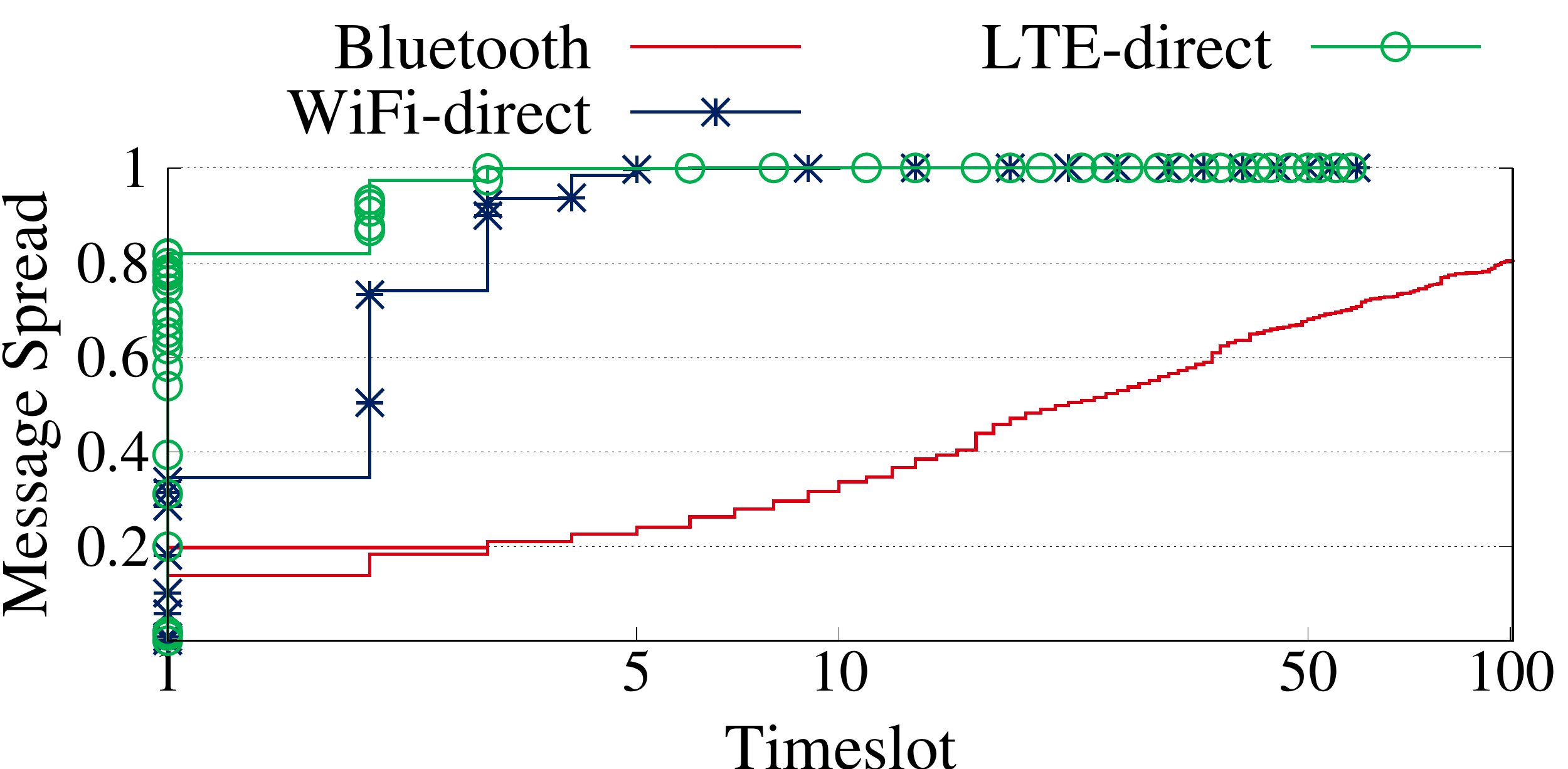}
        \caption{The spread of a message over time.}
        \label{fig:spread}
    \end{subfigure}
    \begin{subfigure}{0.65\columnwidth}
        \centering
        \includegraphics[height=\columnwidth, angle=270]{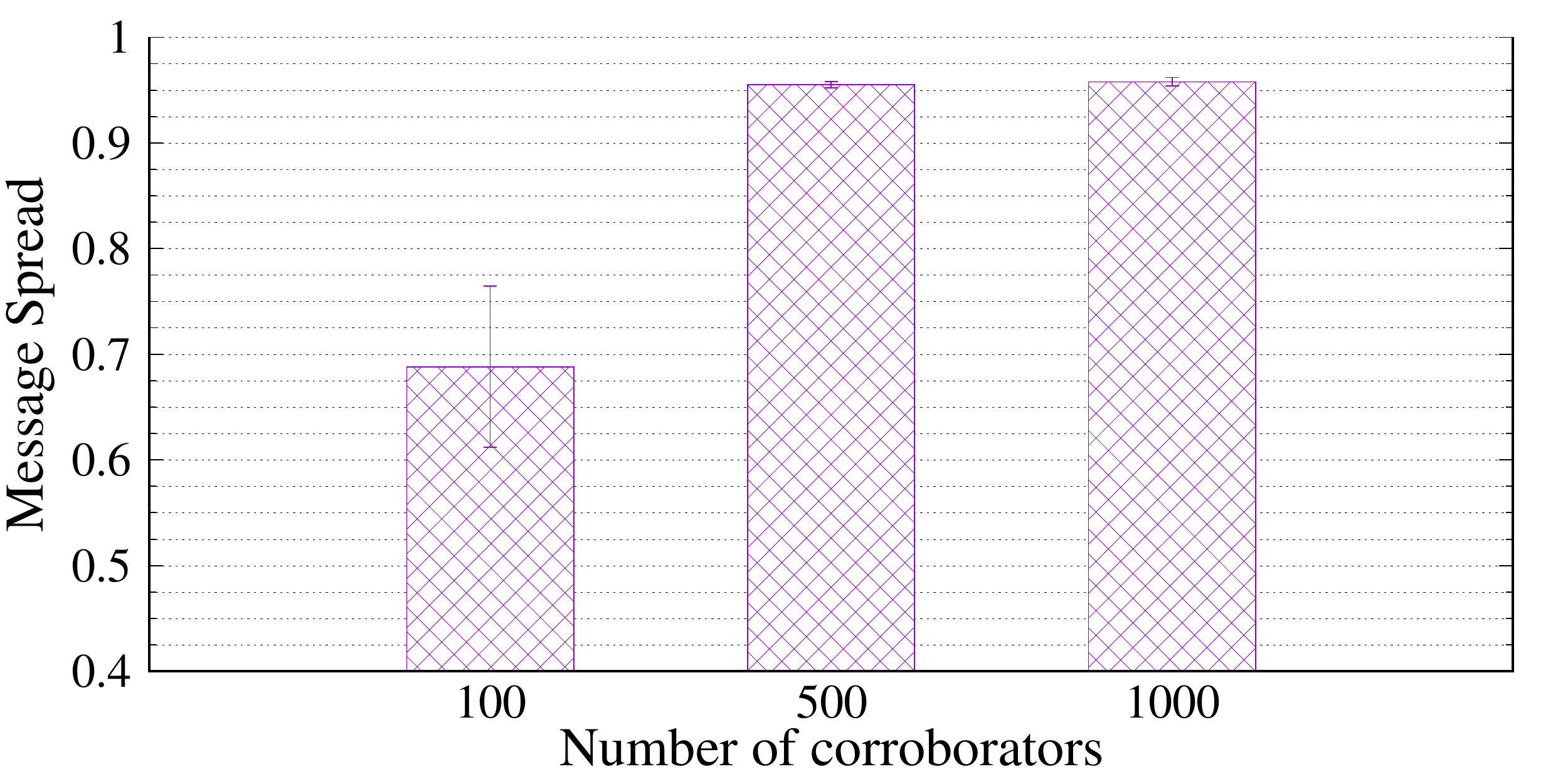}
        \caption{Spread of a message via different numbers of corroborators.}
        \label{fig:spvspop}
    \end{subfigure}
    \begin{subfigure}{0.65\columnwidth}
        \centering
        \includegraphics[height=\columnwidth, angle=270]{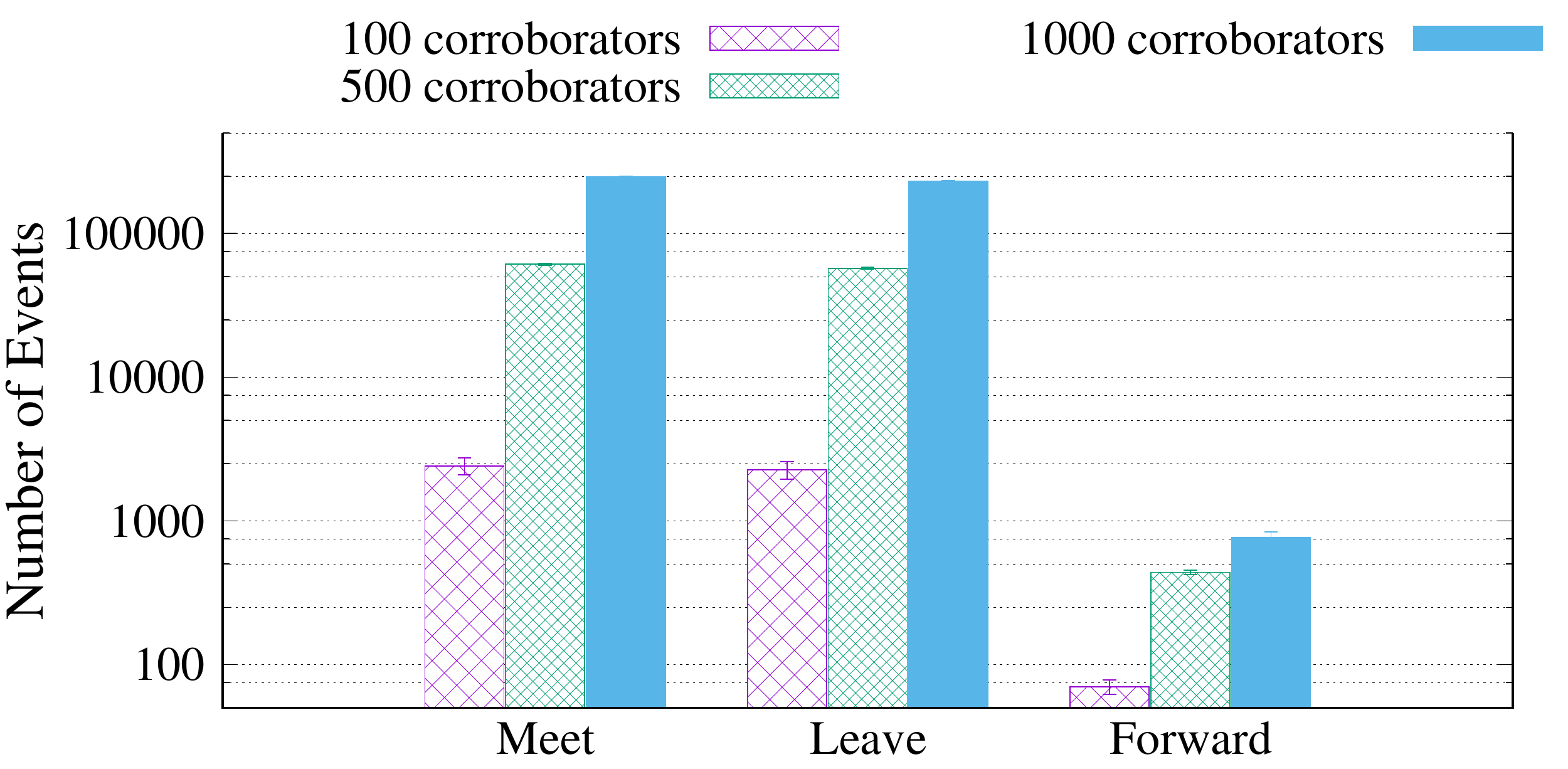}
        \caption{Number of generated events.}
        \label{fig:events}
    \end{subfigure}
    \caption{Experimentation with the parameters of Mneme to characterise the interactions of the corroborators.\vspace{-0.2cm}}
        \label{fig:3}
\end{figure*}
\smallskip\noindent\textbf{Parameters of Mneme.} The operability and robustness of Mneme depends on the size deployment area ($|\mathcal{A}|$), the number of the active corroborators ($\mathcal{N}_a$), the communication technology (e.g., Wifi-direct), the fraction of the malicious users ($f$), and four parameters ($mRS$, $mD$,  $\mathcal{K}_{\tau}$ and $T$).  Mneme has the properties of persistence and liveness if PoC and PoE terminate. PoC and PoE terminate if the network is partially synchronous (i.e., $\Delta < \infty$), which depends on the density of the corroborators in the deployment area and values of the four parameters. In more detail, the higher the density of the of the corroborators (i.e., a high coverage radius of the communication technology, high number of users, small deployment area), the lower the value of $\Delta$, the faster PoC and PoE terminate. The higher the values of $mRS$ and $mD$, the more time PoC needs to terminate and the more difficult it is for a malicious user to double-spend. The higher the value of $\mathcal{K}_{\tau}$ the more time it takes for PoE to terminate and the more difficult it is for a malicious user to claim coins he does not own. The longer the epoch $T$, the slower the storage needs of Mneme will be decreased, and PoE will produce less traffic.

\smallskip\noindent\textbf{Example.} For a coverage radius of 100 meters (which could be achieved by LTE-direct), we find that having more than 500 users per sq km guarantees that any communication would reach all nodes in 10 minutes with a probability $>99\%$. For comparison, the average population density of an urban settlement is typically more than 10,000 people per square kilometer.

\begin{figure*}
\centering
    \begin{subfigure}{0.65\columnwidth}
        \centering
        \includegraphics[height=\columnwidth, angle=270]{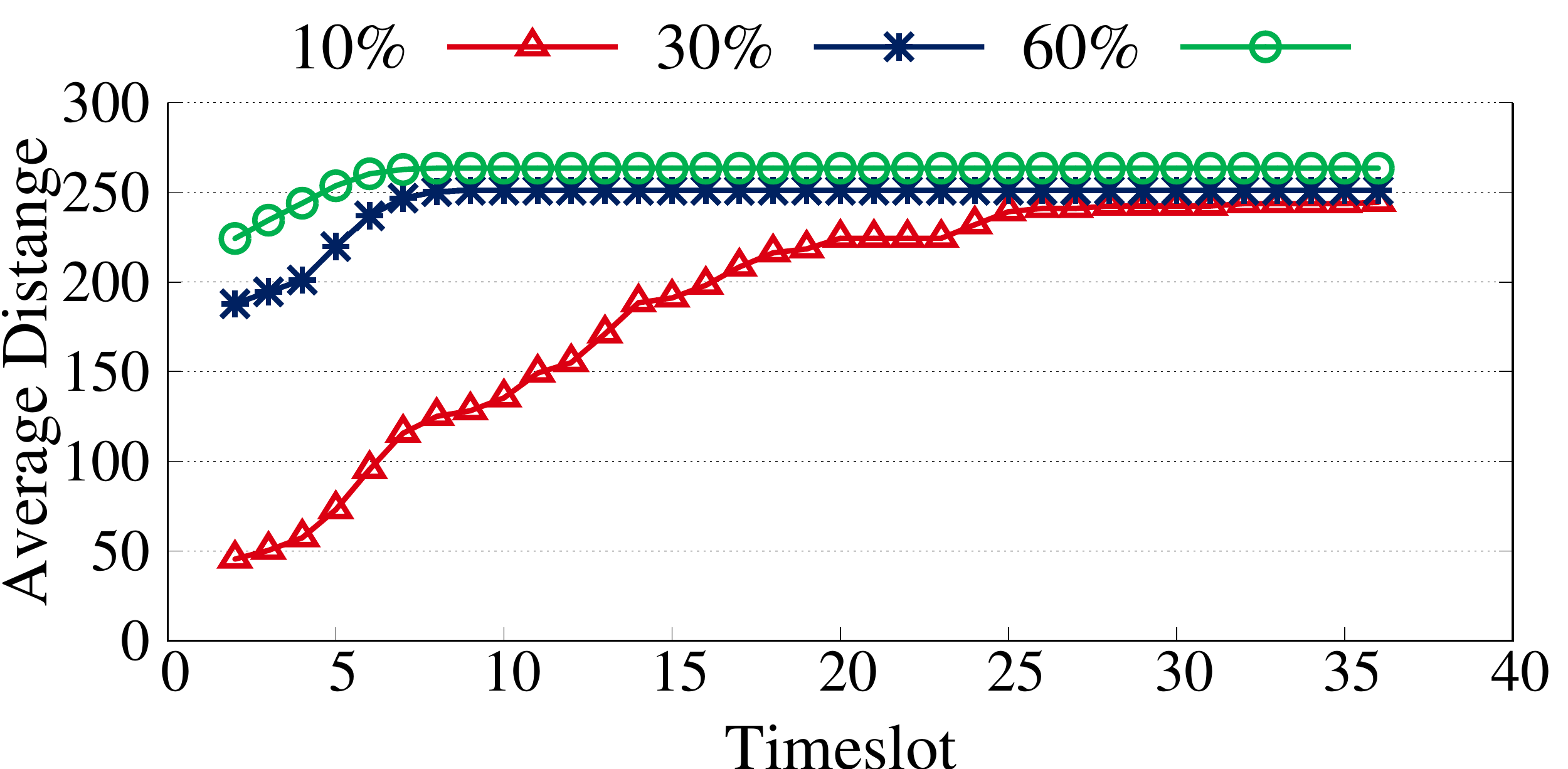}
        \caption{Average distance between the corroborators that can sign a new block over time.}
        \label{fig:avdT}
    \end{subfigure}
    \begin{subfigure}{0.65\columnwidth}
        \centering
        \includegraphics[height=\columnwidth, angle=270]{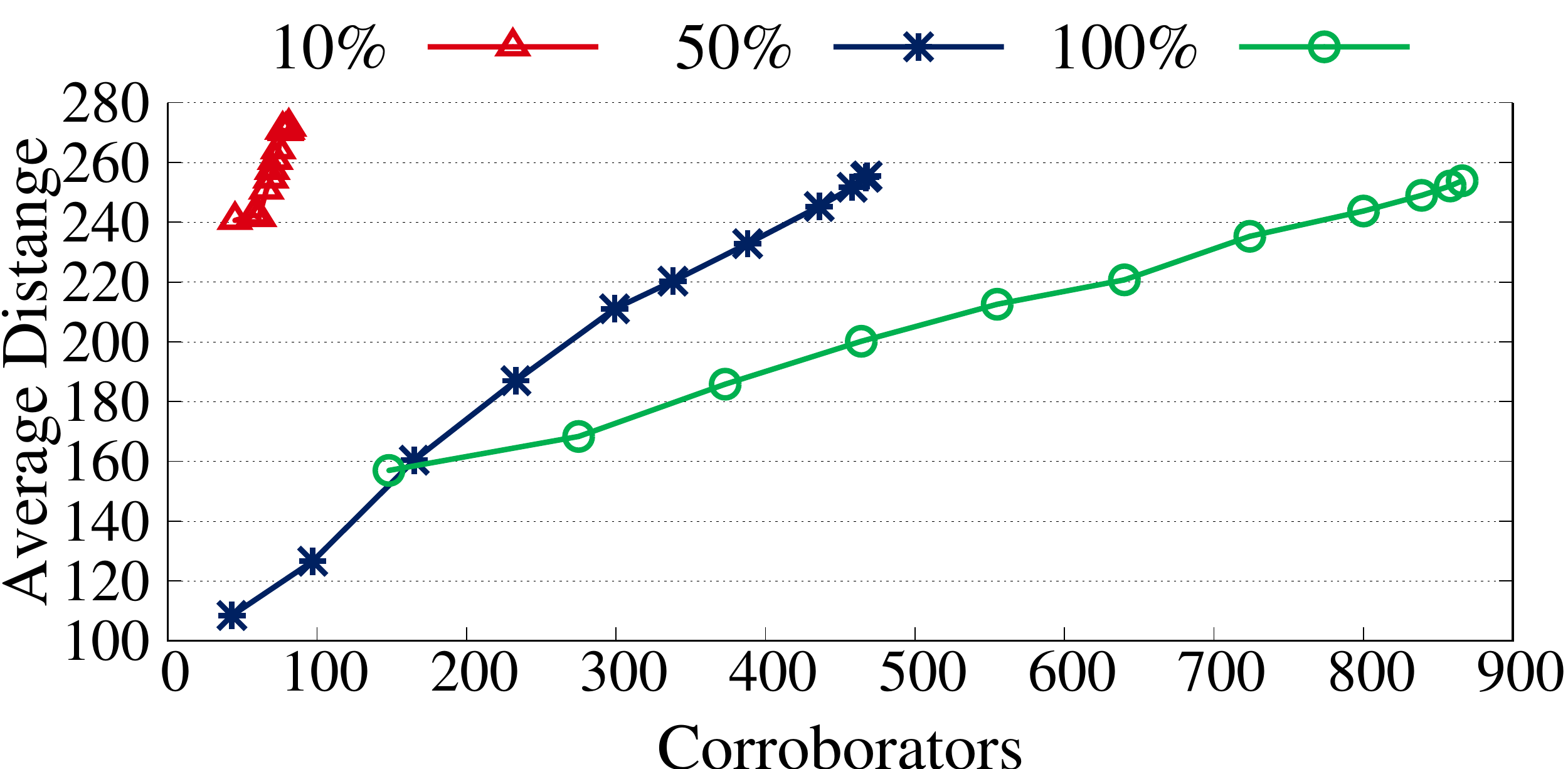}
        \caption{Average distance between the corroborators that can sign a new block for different numbers of corroborators.}
        \label{fig:avdC}
        \end{subfigure}
        \begin{subfigure}{0.65\columnwidth}
            \centering
            \includegraphics[height=\columnwidth, angle=270]{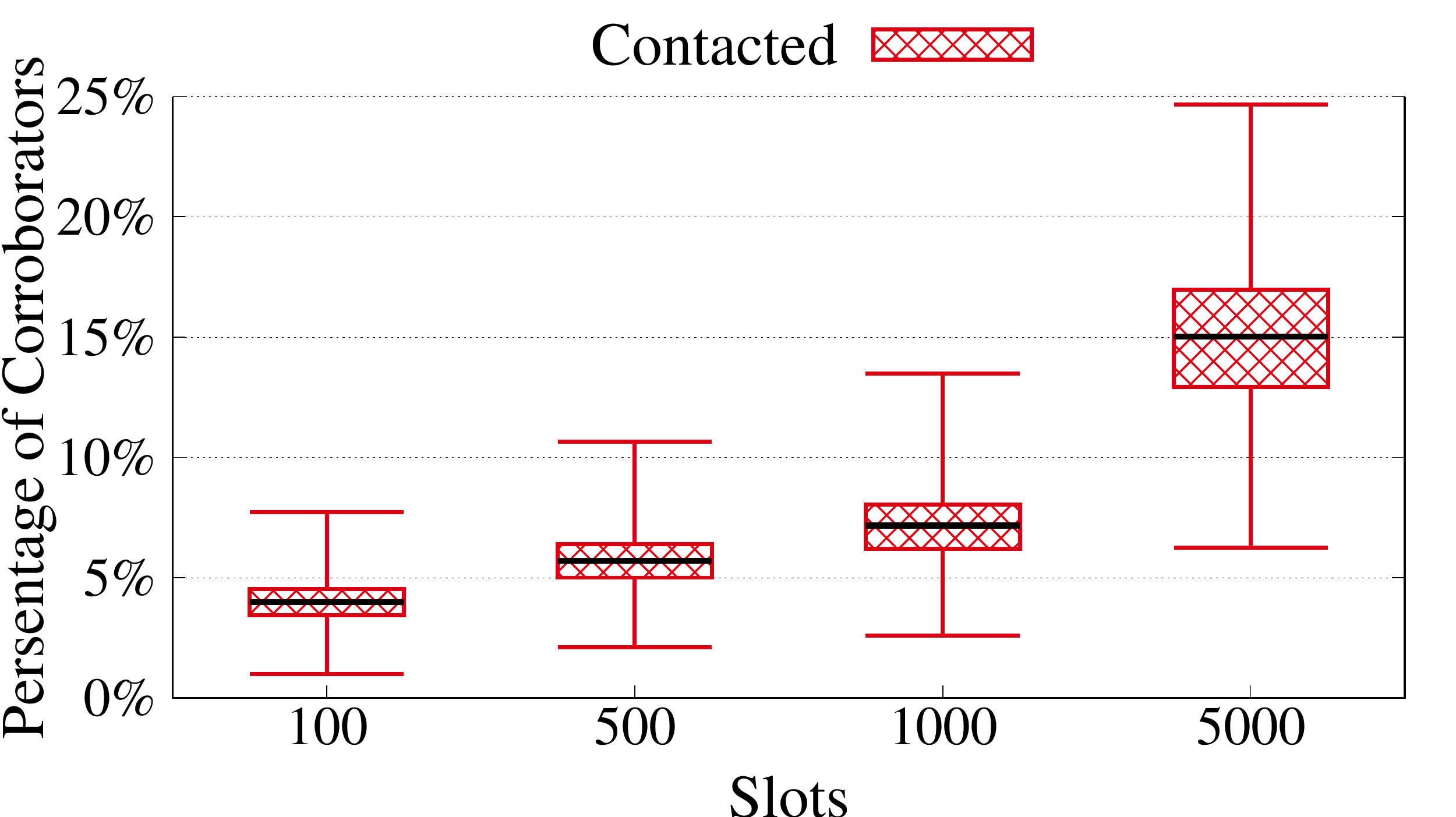}
            \caption{Distribution of the unique meets of corroborators.}
            \label{fig:uniqueMeets}
        \end{subfigure}
    \caption{Impact of the parameters of Mneme in the termination of PoC and PoE consensus protocols.\vspace{-0.2cm}}
    \label{fig:4}
\end{figure*}

\section{Evaluation}\label{sec:eval}
In order to depict the conditions under which PoC and PoE terminate and show that Mneme has the properties of persistence and liveness, we implement an event-driven simulator in JAVA. The input to the simulator includes: the size the deployment area, the number of the mobile users, the employed communication technology, the duration of the simulation and the fraction of the malicious users. We consider an area of 500 meters by 500 meters where users are moving randomly towards any possible direction. We do not consider special mobility patterns or limitations on the users movement and we simulate mobile users who are moving with the same speed (1 meter per slot). We use three communication technologies: Bluetooth, WiFi-direct and LTE-direct. We assume that two users can exchange a message using Bluetooth if the distance between them is less than 20 meters, using WiFi-direct if it is less than 50 and using LTE-direct if it is less than 100. If two users can exchange a message the simulator produces a ``MEET" event and when they loose this ability it produces a ``LEAVE"~event. 

\smallskip\noindent\textbf{Network Characterisation.}
Partially synchronous networks are characterised by a parameter $\Delta$ that represents an upper bound on the time needed for a message to be transferred to every active user\footnote{In synchronous networks, all the users are active all the time while in asynchronous networks $\Delta = \infty$.}. We analysed three scenarios with 1000 users in 100 slots and three communication technologies to measure $\Delta$ since it is directly related to the termination of PoC and PoE. Figures~\ref{fig:spread},~\ref{fig:spvspop} and~\ref{fig:events} show the results. Since LTE-direct has higher coverage radius, the messages are spread faster while in the case of Bluetooth time needed for a message to reach every corroborator is much higher. We also examine the impact of the number of users in the spread of a message in 100 slots and when the users communicate via WiFi-direct. Figure~\ref{fig:spvspop} shows how the message is spread in the cases of 100, 500 and 1000 users. 500 and 1000 users can guarantee the delivery of the message to every user while in the case of 100 users, 100 slots are not enough since less than 70\% of the corroborators receive the message. This fact is also verifiable by the relatively high standard deviation of the fraction of the users who have received the message. Figure~\ref{fig:events} depicts the number of the produced events (``MEET", ``LEAVE" and ``FORWARD") when a message is spread using WiFi-direct for different number of users.  The more the users, the more events are produced. So, the higher the coverage radius of the communication technology, the faster PoC and PoE will terminate and the more the active corroborators the higher the probability for PoC and PoE to terminate within a given~period. 

\smallskip\noindent\textbf{Approximating $\Delta$.} 
\label{approximate_procedure}
Since $\Delta$ is not in common knowledge of every user, each user would need to calculate the approximate upper bound $\Delta$ by itself. The $\Delta$ depends upon two parameters: (a) The location of the user which he can determine using GPS (Figure \mbox{\ref{fig:rgg_3d}} and \mbox{\ref{fig:rgg_3d_2}} demonstrate the variation of $\Delta$ with the location of the user) (b) The network characteristics which may vary with time and cannot be determined deterministically. We describe a procedure to approximate $\Delta$ as follows: 
\begin{enumerate}
    \item The user sends a message containing a timestamp ($a_i$) to some of his trusted users. 
    \item The trusted users reply with the message along with the timestamp of when they received the message ($b_i$), timestamp of when they send their reply($c_i$) and their location $(x_i, y_i)$.
    \item The user calculates the distance between the trusted user and himself as $d_i = \sqrt{(x_i - x_o)^2+(y_i - y_o)^2}$ and the time average time difference as $t_i = \frac{(b_i-a_i) + (d_i-c_i)}{2}$.
    \item Let matrix $A = \begin{bmatrix}
    d_1 &1\\
    d_2 &1\\
    \ldots &\ldots\\
    \end{bmatrix}$, $x = \begin{pmatrix}p q\end{pmatrix}$, $y = \begin{bmatrix}
    t_1\\
    t_2\\
    \ldots
    \end{bmatrix}$
    
    \item Let $x = (A^TA)^{-1}A^Ty$ then we can set \[\Delta = p\times(\max(x_o^2,(1-x_o)^2)+\max(y_o^2,(1-y_o^2))+q\] 
\end{enumerate}
\smallskip\noindent\textbf{Friendly Relaxation.}
Since Mneme would be deployed in a local setting where all users are in vicinity of each other, it is likely that users know each other socially as well. In that case, if the recipient trusts the sender of the transaction that he/she will not double spend or the transaction is of nominal value, the recipient could skip or reduce $\delta$ for the particular transaction at his/her convenience to accommodate quicker commerce.


\smallskip\noindent\textbf{Termination of PoC.}
We measure the average distance between the users that can verify a block for different values of $\rho_{b}^{u}$ to examine whether PoC terminates.  Figure~\ref{fig:avdT} shows how the average distance between the users increases over time for three different values of $\rho_{b}^{u}$, 10\%, 30\% and 60\%. Figure~\ref{fig:avdC} depicts how the increase in $\rho_{b}^{u}$ slows down the increase of the average distance between the corroborators. This is expected since the more probable it is for a corroborator to sign a block, the more corroborators will be found close to the one that initiated PoC and the more extra corroborators will be needed to reach an average distance between them, which is higher than mD. This result combined with the results of the previous paragraph shows that PoC terminates if mRS and mD are correctly selected (e.g., in the examined setting PoC will not terminate if mD is 300 meters or more). 

\begin{figure}[t]
        \centering
        \includegraphics[height=0.7\columnwidth, angle=270]{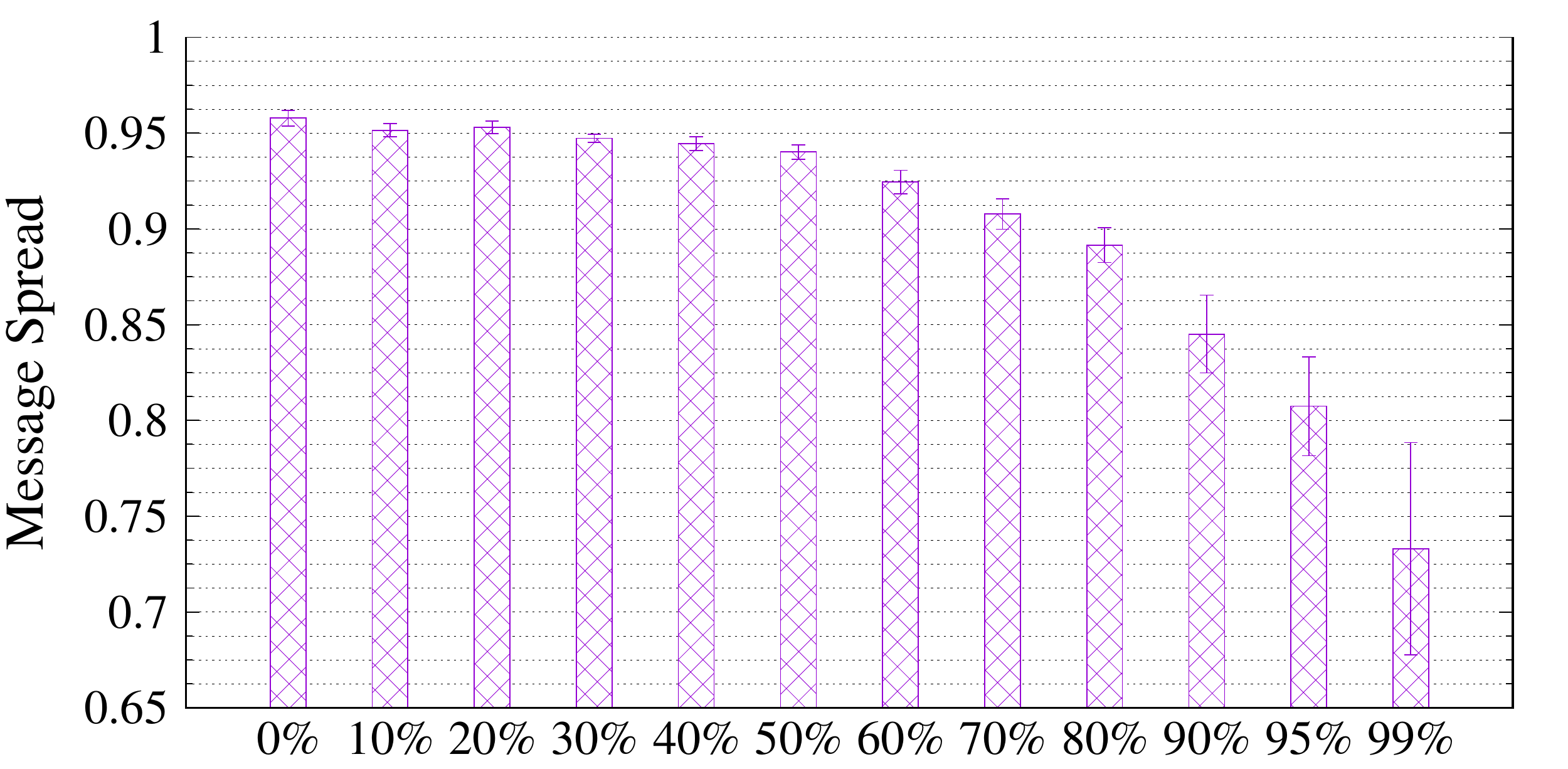}
        \caption{Impact of non-assisting users to forwarding.\vspace{-0.2cm}}
        \label{fig:imofMal}
\end{figure}

\smallskip\noindent\textbf{Termination of PoE.} We measure what fraction of the total users each user meets  to examine whether PoE terminates. By measuring that quantity, we can estimate $\theta_{\tau}^{u}$ and determine the minimum number of corroborators needed to execute PoE. Figure~\ref{fig:uniqueMeets} depicts four scenarios with 1000 users and four values of simulation duration. Each experiment is depicted via a candlestick where the black line shows the average, the lower line the minimum, the upper line the maximum and the boxed area contains the values between the 25$^{th}$ percentile and 75$^{th}$ percentile. Using Figure~\ref{fig:uniqueMeets} we can argue that if $\mathcal{K}_{\tau}^{m}$ is less than 5\% of the total population and the duration of $\tau$ is 500 slots, PoE terminates with probability 75\% but if the slots are 5000, PoE terminates with certainty. Next, in Figure~\ref{fig:imofMal} we examine the impact of users who are not assisting in the delivery of messages. Via this plot, we can see that every message has more than 70\% chances to be delivered even if 99\% of the users are not helping with forwarding. Figures~\ref{fig:uniqueMeets} and \ref{fig:imofMal} show that PoE will terminate if the epoch duration and the number of the selected users are carefully selected. 

Given Figures~\ref{fig:3},~\ref{fig:4} and~\ref{fig:imofMal}, we can infer that PoC and PoE terminate when \textit{(i)} there are enough users to maintain Mneme, \textit{(ii)} mRS and mD are carefully selected to guarantee the termination of PoC and \textit{(iii)} the minimum number of corroborators needed to execute PoE is high enough and the epoch duration of PoE is long enough to guarantee security. 

\section{Discussion}\label{sec:disc}


In this work, we implement the scheme proposed in Localcoin to replace the {\mbox{\it computational}} hardness that is at the root of Bitcoin's security with the {\mbox{\it social}} hardness of ensuring that all witnesses to a transaction are colluders (users assisting the malicious user to double spend)~\cite{2017arXiv170808086C}. Our design requires users to sacrifice anonymity to maintain reputation scores, which is a threat to privacy due to the fact that anyone can read its content of the blockchain, trace transactions and perform deanonymization attacks. Especially on the case we are focusing on, where corroborators are employing their context to verify blocks and maintain the ledger, their identity is more exposed. Unfortunately, most of the existing proposed solutions are based on zero-knowledge proofs that are computationally expensive~\cite{ben2014zerocash} 
The interconnectivity of the mobile users and their mobility patterns cannot be easily predicted by others, and this results in increases in traffic and energy consumption of the devices~\cite{song2010limits}. 
The performance of protocols that maintain a distributed ledger depends on energy-efficient neighbor discovery protocols that are complemented with bookkeeping functionalities that can be used on mobility prediction. Incentives are of high importance in the performance of Mneme. We introduced four types of fees to motivate corroborators to contribute resources by forwarding transactions and block messages, and storing transactions and blocks on their devices. However, a detailed analysis of the incentives is left for future work


Whenever Mnene is employed to non-financial applications, the exchanged transactions transfer data and fees. The fees have a central role since they are needed to motivate the corroborators to share their resources. In order to enable the merging of non-financial transactions, each transaction includes a field that indicates a function over the data and the number of the occurences. PoE will be based on this field. For example, if one mobile user created three transactions to transfer a sensor reading to another mobile user and the function over the data is the average, PoE can produce a new transaction with the average data value and set the occurrences field to three. 

\section{Conclusion}\label{sec:concl}
Distributed ledgers that are maintained solely by mobile devices can be employed by several mobile applications. The two most popular distributed ledgers are blockchain and directed acyclic graph. Both of them employ protocols that cannot function on mobile devices because they are designed for computationally powerful and well-connected nodes. 
In this work, we identified the need for a distributed ledger that can be maintained solely by mobile devices and designed Mneme to fill this need. Also we introduced two novel consensus protocols that are based on the users' context and serviceableness. The first one is responsible for data insertion in Mneme and the second one deletes blocks from Mneme to release storage resources. We prove that both protocols are secure against popular attacks and can guarantee the operability of Mneme. Experimental results provide a deeper understanding of the parameters of Mneme and the trade-off between performance and security. Additionally, advancing on~\cite{9155497} we extended the adversarial analysis and provided a detailed analysis against double spending attacks.




\bibliography{bibliography}
\bibliographystyle{IEEEtran}

\begin{IEEEbiography}[{\includegraphics[width=1in,height=1.25in,clip,keepaspectratio]{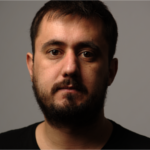}}]{Dimitris Chatzopoulos} is a research assistant professor at The Hong Kong University of Science and Technology (HKUST). Dimitris
received his PhD in Computer Science and Engineering from HKUST and his Diploma and Msc in Computer Engineering and Communications from the Department of Electrical and Computer Engineering of University of Thessaly, Volos, Greece. His main research interests are in the areas of mobile computing, device--to--device ecosystems and cryptocurrencies.
\end{IEEEbiography}

\vspace{-1cm}

\begin{IEEEbiography}[{\includegraphics[width=1in,height=1.25in,clip,keepaspectratio]{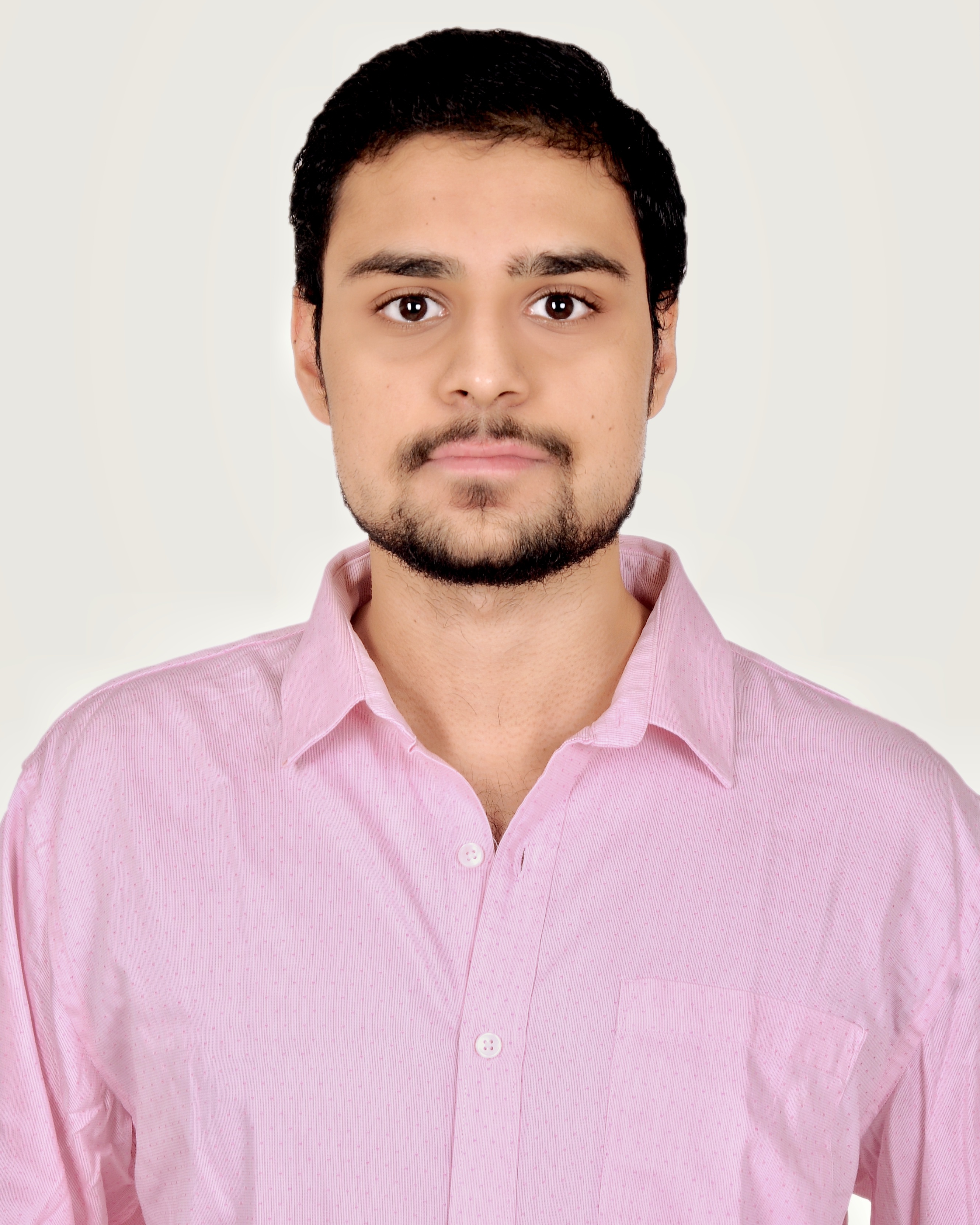}}]{Anurag Jain} is an undergraduate researcher at the Machine Learning Laboratory@IIITH. As a young researcher, he has diverse interests including Blockchains, Game Theory, Mechanism Design and Economics. He has also co-authored 2 papers in the workshop on Game Theory in Blockchain at WINE 2020.
\end{IEEEbiography}

\vspace{-1.1cm}

\begin{IEEEbiography}[{\includegraphics[width=1in,height=1.25in,clip,keepaspectratio]{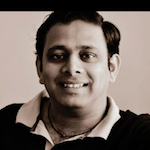}}]{Sujit Gujar} is an Assistant Professor at the Machine Learning Laboratory@IIITH. Prior to this, he was a Sr. Research Associate at Indian Institute of Science. He worked as a post-doctoral researcher at Ecole polytechnique federale de Lausanne (EPFL). He also worked as a research scientist with Xerox Research Centre India where he contributed in developing a technology that enables enterprises to use crowdsourcing as a complimentary workforce. His research interests are Game Theory, Mechanism Design, Machine Learning, and Cryptography applied to modern web and AI applications such as Auctions, Internet Advertising, Crowdsourcing, and multi-agent systems. His doctoral thesis was awarded alumni medal for best doctoral thesis in the Department of Computer Science and Automation at Indian Institute of Science. He was a recipient of Infosys fellowship for his doctoral research. He has co-authored 7 journal publications, 1 book chapter and 50 peer reviewed international conference/workshop papers. He has 11 patents on his name.
\end{IEEEbiography}

\vspace{-1.1cm}

\begin{IEEEbiography}[{\includegraphics[width=1in,height=1.25in,clip,keepaspectratio]{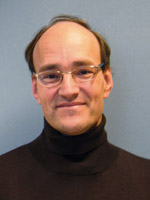}}]{Boi Faltings} is a full professor of computer science at the Ecole Polytechnique Federale de Lausanne (EPFL), where he heads the Artificial Intelligence Laboratory, and has held visiting positions at NEC Research Institute, Stanford University and the HongKong University of Science and Technology. He has co-founded 6 companies in e-commerce and computer security and acted as advisor to several other companies. Prof. Faltings has published over 300 refereed papers and graduated over 30 Ph.D. students, several of which have won national and international awards. He is a fellow of the European Coordinating Committee for Artificial Intelligence and a fellow of the Association for Advancement of Artificial Intelligence (AAAI). He holds a Diploma from ETH Zurich and a Ph.D. from the University of Illinois at Urbana-Champaign.
 
\end{IEEEbiography}

\vspace{-1.3cm}

\begin{IEEEbiography}[{\includegraphics[width=1in,height=1.25in,clip,keepaspectratio]{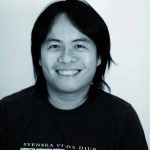}}]{Pan Hui} (IEEE Fellow and ACM Distinguished Scientist)
received his Ph.D degree from Computer Laboratory, University of Cambridge, and earned his MPhil and BEng both from the Department of Electrical and Electronic Engineering, University of Hong Kong. He is currently a faculty member of the Department of Computer Science and Engineering at the Hong Kong University of Science and Technology where he directs the HKUST-DT System and Media Lab. He also serves as a Distinguished Scientist of Telekom Innovation Laboratories (T-labs) Germany and an adjunct Professor of social computing and networking at Aalto University Finland.  
\end{IEEEbiography}



\end{document}